\documentclass[longauth]{aa}  

\usepackage{graphicx,subcaption}
\usepackage{txfonts}
\usepackage[switch]{lineno}
\usepackage{natbib}
\usepackage{lscape}
\usepackage{longtable}
\bibpunct{(}{)}{;}{a}{,}{,}

\usepackage{graphicx}
\usepackage{pifont}

\usepackage{pstricks}
\usepackage{color}
\graphicspath{ {figures/} }

\usepackage{hyperref}
\hypersetup{
    bookmarks=true,         
    unicode=true,           
    pdftoolbar=true,        
    pdfmenubar=true,        
    pdffitwindow=true,      
    pdftitle={Characterization of asteroid (41) Daphne and its satellite},
    pdfauthor={Carry et al.},
    pdfsubject={Planetary Science},
    pdfkeywords={},         
    pdfnewwindow=true,      
    colorlinks=true,        
    linkcolor=gray,         
    citecolor=blue,         
    filecolor=gray,         
    urlcolor=gray           
}

\newcommand{\add}[1]{#1}
\newcommand{\rem}[1]{}

\newcommand{\numb}[1]{#1}

\newcommand{\koala}{\texttt{KOALA}}
\newcommand{\adam}{\texttt{ADAM}}
\newcommand{\genoid}{\texttt{Genoid}}
\newcommand{\mistral}{\texttt{Mistral}}

\newcommand{\SatOne}{\textsl{S/2008 (41) 1}}
\newcommand{\NameOne}{\textsl{Peneius}}
\newcommand{\ObsOne}{30}

\newcommand{\rmsOne}{8.3}
\newcommand{\DMag}{9.49\,$\pm$\,0.32}

\newcommand{\obsDay}{3783}
\newcommand{\obsRev}{3326}
\newcommand{\orbPer}{1.137\,446\,$\pm$\,0.000\,009}

\newcommand{\sLon}{199}
\newcommand{\sLat}{-33}
\newcommand{\sPer}{5.98798}

\newcommand{\Diam}{187}
\newcommand{\dDiam}{21.5}

\newcommand{\satDiam}{2.4$_{-1.1}^{+1.8}$}

\newcommand{\Dens}{1.77}
\newcommand{\dDens}{0.26}

\newcommand{\Poro}{17}
\newcommand{\dPoro}{3}

\newcommand{\sid}{g$\cdot$cm$^{-3}$}

\begin{document} 

  \title{The homogeneous internal structure of\\ CM-like asteroid (41) Daphne\thanks{Based on observations made with 
      1) ESO Telescopes at the La Silla
      Paranal Observatory under programs
      \href{http://archive.eso.org/wdb/wdb/eso/abstract/query?ID=8150110}{281.C-5011}
      (PI Dumas),
      \href{http://archive.eso.org/wdb/wdb/eso/abstract/query?ID=9900980}{099.D-0098}
      (SPHERE GTO),
      and
      \href{http://archive.eso.org/wdb/wdb/eso/abstract/query?ID=9900740&progid=199.C-0074(A)}{199.C-0074(A)}
      (PI Vernazza); and
      2) the W. M. Keck Observatory, which is operated as a scientific
      partnership among the California Institute of Technology, the
      University of California and the National Aeronautics and Space
      Administration. The Observatory
      was made possible by the generous financial support of the W. M. Keck
      Foundation.},
    \thanks{The reduced and deconvolved AO images and the 3-D shape
      model are publicly available at
      \href{http://observations.lam.fr/astero/}{http://observations.lam.fr/astero/}
    }
  }

  \author{B.~Carry\inst{\ref{oca}}         \and 
    F.~Vachier\inst{\ref{imcce}}           \and 
    J.~Berthier\inst{\ref{imcce}}          \and 
    M.~Marsset\inst{\ref{qub}}             \and 
    P.~Vernazza\inst{\ref{lam}}            \and 
    J.~Grice\inst{\ref{oca},\ref{ou}}      \and 
    W.~J.~Merline\inst{\ref{swri}}         \and 
    E.~Lagadec\inst{\ref{oca}}             \and 
    A.~Fienga\inst{\ref{geoazur}}          \and 
    A.~Conrad\inst{\ref{lbt}}              \and 
    E.~Podlewska-Gaca\inst{\ref{poznan},\ref{edyta}} \and 
    T.~Santana-Ros\inst{\ref{poznan}}      \and 
    M.~Viikinkoski\inst{\ref{tampere}}     \and 
    J.~Hanu{\v s}\inst{\ref{prague}}       \and 
    C.~Dumas\inst{\ref{tmt}}               \and 
    J.~D.~Drummond\inst{\ref{afrl}}        \and 
    P.~M.~Tamblyn\inst{\ref{swri},\ref{pt}}\and 
    C.~R.~Chapman\inst{\ref{swri}}         \and 
    R.~Behrend\inst{\ref{cdr}}             \and 
    L.~Bernasconi\inst{\ref{cdr}}          \and 
    P.~Bartczak\inst{\ref{poznan}}         \and 
    Z.~Benkhaldoun\inst{\ref{ouk}}         \and 
    M.~Birlan\inst{\ref{imcce}}            \and 
    J.~Castillo-Rogez\inst{\ref{jpl}}      \and 
    F.~Cipriani\inst{\ref{estec}}          \and 
    F.~Colas\inst{\ref{imcce}}             \and 
    A.~Drouard\inst{\ref{lam}}             \and 
    J.~{\v D}urech\inst{\ref{prague}}      \and 
    B.~L.~Enke\inst{\ref{swri}}            \and 
    S.~Fauvaud\inst{\ref{cdr},\ref{bardon}}\and 
    M.~Ferrais\inst{\ref{liege}}           \and 
    R.~Fetick\inst{\ref{lam}}              \and 
    T.~Fusco\inst{\ref{lam}}               \and 
    M.~Gillon\inst{\ref{liege}}            \and 
    E.~Jehin\inst{\ref{liege}}             \and 
    L.~Jorda\inst{\ref{lam}}               \and 
    M.~Kaasalainen\inst{\ref{tampere}}     \and 
    M.~Keppler\inst{\ref{mpia}}            \and 
    A.~Kryszczynska\inst{\ref{poznan}}     \and 
    P.~Lamy\inst{\ref{lam}}                \and 
    F.~Marchis\inst{\ref{seti}}            \and 
    A.~Marciniak\inst{\ref{poznan}}        \and 
    T.~Michalowski\inst{\ref{poznan}}      \and 
    P.~Michel\inst{\ref{oca}}              \and 
    M.~Pajuelo\inst{\ref{imcce},\ref{puc}} \and 
    P.~Tanga\inst{\ref{oca}}               \and 
    A.~Vigan\inst{\ref{lam}}               \and 
    B.~Warner\inst{\ref{warner}}           \and 
    O.~Witasse\inst{\ref{estec}}           \and 
    B.~Yang\inst{\ref{eso}}                \and 
    A.~Zurlo\inst{\ref{lam},\ref{ud1},\ref{ud2}} 
  }

   \institute{
     %
     Universit\'e C{\^o}te d'Azur, Observatoire de la
     C{\^o}te d'Azur, CNRS, Laboratoire Lagrange, France
     \email{benoit.carry@oca.eu}\label{oca}
     \and 
     IMCCE, Observatoire de Paris, PSL Research University, CNRS, Sorbonne Universit{\'e}s, UPMC Univ Paris 06, Univ. Lille, France
     \label{imcce}
     \and 
     Astrophysics Research Centre, Queen's University Belfast, BT7 1NN, UK
     \label{qub}
     \and 
     Aix Marseille Univ, CNRS, LAM, Laboratoire d'Astrophysique de Marseille, Marseille, France
     \label{lam}
     \and 
     Open University, School of Physical Sciences, The Open University, MK7 6AA, UK
     \label{ou}
     \and 
     Southwest Research Institute, Boulder, CO 80302, USA
     \label{swri}
     \and 
     Universit\'e C{\^o}te d'Azur, Observatoire de la
     C{\^o}te d'Azur, CNRS, Laboratoire G{\'e}oAzur, France
     \label{geoazur}
     \and 
     Large Binocular Telescope Observatory, University of Arizona, Tucson, AZ 85721, USA
     \label{lbt}
     \and 
     Faculty of Physics, Astronomical Observatory Institute, Adam Mickiewicz University, ul. S{\l}oneczna 36, 60-286 Pozna{\'n}, Poland
     \label{poznan}
     \and 
     Oukaimeden Observatory, High Energy Physics and Astrophysics Laboratory, Cadi Ayyad University, Marrakech, Morocco
     \label{ouk}
     \and 
     Institute of Physics, University of Szczecin, Wielkopolska 15, 70-453 Szczecin, Poland
     \label{edyta}
     \and 
     Department of Mathematics, Tampere University of Technology, PO Box 553, 33101, Tampere, Finland
     \label{tampere}
     \and 
     Max Planck Institute for Astronomy, K{\"o}nigstuhl 17, D-69117, Heidelberg, Germany 
     \label{mpia}
     \and 
     Astronomical Institute, Faculty of Mathematics and Physics, Charles University, V~Hole{\v s}ovi{\v c}k{\'a}ch 2, 18000 Prague, Czech Republic
     \label{prague}
     \and 
     Thirty-Meter-Telescope, 100 West Walnut St, Suite 300, Pasadena, CA 91124, USA
     \label{tmt}
     \and 
     Leidos, Starfire Optical Range, AFRL, Kirtland AFB, NM 87117, USA
     \label{afrl}
     \and 
     Binary Astronomy, Aurora, CO 80012, USA
     \label{pt}
     \and 
     CdR \& CdL Group: Lightcurves of Minor Planets and Variable Stars, Observatoire de Gen{\`e}ve, CH-1290 Sauverny, Switzerland
     \label{cdr}
     \and 
     Jet Propulsion Laboratory, California Institute of Technology, 4800 Oak Grove Drive, Pasadena, CA 91109, USA
     \label{jpl}
     \and 
     European Space Agency, ESTEC - Scientific Support Office, Keplerlaan 1, Noordwijk 2200 AG, The Netherlands
     \label{estec}
     \and 
     Observatoire du Bois de Bardon, 16110 Taponnat, France
     \label{bardon}
     \and 
     Space sciences, Technologies and Astrophysics Research (STAR) Institute, Universit{\'e} de Li{\`e}ge, All{\'e}e du 6 Ao{\^u}t 17, 4000 Li{\`e}ge, Belgium
     \label{liege}
     \and 
     SETI Institute, Carl Sagan Center, 189 Bernado Avenue, Mountain View CA 94043, USA 
     \label{seti}
     \and 
     Secci{\'o}n F{\'i}sica, Departamento de Ciencias, Pontificia Universidad Cat{\'o}lica del Per{\'u}, Apartado 1761, Lima, Per{\'u}
     \label{puc}
     \and 
     Center for Solar System Studies, 446 Sycamore Ave., Eaton, CO 80615, USA
     \label{warner}
     \and 
     European Southern Observatory (ESO), Alonso de Cordova 3107, 1900 Casilla Vitacura, Santiago, Chile
     \label{eso}
     \and 
     N{\'u}cleo de Astronomía, Facultad de Ingenier{\'i}a y Ciencias,
     Universidad Diego Portales, Av. Ejercito 441, Santiago, Chile
     \label{ud1}
     \and 
     Escuela de Ingeniería Industrial, Facultad de Ingenier{\'i}a y Ciencias,
     Universidad Diego Portales, Av. Ejercito 441, Santiago, Chile
     \label{ud2}
   }


   \date{Received 2018; accepted 2019-01-07}

\pagebreak 
  \abstract
   {\add{CM-like asteroids (Ch and Cgh classes)
    are a major population within the broader C-complex, encompassing
    about 10\% of the mass of the main asteroid belt.
    Their internal structure has been predicted to be homogeneous,
    based on their compositional similarity as inferred from spectroscopy
     (Vernazza et al., 2016, AJ 152, 154)
     and numerical modeling of their early thermal evolution
     (Bland \& Travis, 2017, Sci. Adv. 3, e1602514).
   }}
   {
     \add{Here we aim to test this hypothesis by
     deriving the density of the CM-like asteroid (41) Daphne from
     detailed modeling of its shape and the orbit of its
     small satellite. }
   }
   {\add{We observed Daphne and its satellite within our imaging survey with the Very
     Large Telescope extreme adaptive-optics SPHERE/ZIMPOL camera
     (ID \href{http://archive.eso.org/wdb/wdb/eso/abstract/query?ID=9900740&progid=199.C-0074(A)}{199.C-0074},
     PI P. Vernazza) and complemented this data set with earlier
     Keck/NIRC2 and VLT/NACO observations.}
     We analyzed the dynamics of the satellite with our
     \genoid~meta-heuristic algorithm. Combining our high-angular
     resolution images with optical lightcurves and stellar occultations, we
     determine the spin period, orientation, and 3-D shape, using
     our \adam~shape modeling algorithm.
   }
   {The satellite orbits Daphne on an equatorial, quasi-circular,
     prograde orbit, like the satellites of many other large
     main-belt asteroids.
     The shape model of Daphne reveals several \add{large flat areas that could be
     large impact craters.}
     The mass determined from this orbit
     combined with the volume computed from the shape model implies a
     \add{density for Daphne of \numb{\Dens\,$\pm$\dDens}\,\sid{} (3\,$\sigma$).
     This density is consistent with a primordial CM-like homogeneous
     internal structure with some level of macroporosity ($\approx$\numb{\Poro}\%).}
} 
   {Based on our analysis of the density of Daphne and \numb{75} other 
     Ch/Cgh-type asteroids gathered from the literature, \add{we conclude
     that the primordial internal structure of the CM parent bodies was
     homogeneous.} }


   \maketitle
%
\section{Introduction}
%

  \indent \add{The C-complex encompasses 50\% of the mass of the
  asteroid belt
  \citep[or 14\% if the four largest bodies,
    Ceres, Vesta, Pallas and Hygeia,
    are disregarded,][]{2013-Icarus-226-DeMeo, 2014-Nature-505-DeMeo}.
  Within this complex, the Ch- and Cgh-types are defined
  by the presence of an absorption band around 0.7\,$\mu$m, and a
  UV-dropoff (sharper for the Cgh).
  These have been estimated to represent between 30\%
  and 65\% by number \citep{2012-Icarus-221-Rivkin,
    2014-Icarus-233-Fornasier}, and are
  associated with CM chondrites 
  \citep[namely the Ch- and Cgh-types, see][]{1993-Icarus-105-Vilas, 1998-MPS-33-Burbine,
    2002-Icarus-158-BusII, 2009-Icarus-202-DeMeo,
    2013-AA-554-Lantz, 2014-Icarus-233-Fornasier,
    2016-AJ-152-Vernazza, 2017-MPS-48-Takir}.
  In other words, CM-like bodies represent a significant fraction of
  the C-complex population and encompass
  about 10\% of the mass of all main-belt asteroids. They
  are spread over the entire Main Belt, and are found at all diameters
  \citep[e.g.,][]{2012-Icarus-221-Rivkin, 2014-Icarus-233-Fornasier}.}

  \indent Their absorption band at 0.7\,$\mu$m has been associated
  with phyllosilicates
  \citep{1996-Icarus-124-Vilas}. \add{This is supported 
  by the presence of a phyllosilicate band near 2.8\,$\mu$m
  \citep[commonly called the 3\,$\mu$m
    absorption,][]{2015-AJ-150-Rivkin},
  as well as a shallow band at 2.33\,$\mu$m.
  The 2.8\,$\mu$m band is very similar to that seen in the B-type
  asteroid (2) Pallas 
  \citep{2012-Icarus-219-Takir, 2015-AJ-150-Rivkin} and
  interpreted as being due to serpentine  
  \citep[e.g.,][]{2017-MPS-48-Takir}.
  The 2.33\,$\mu$m band is also associated with the serpentine group
  \citep[i.e., hydrous magnesium-iron phyllosilicates,][]{2018-Icarus-313-Beck}}.

  \indent \add{The meteorites originating from these bodies, the CM chondrites,}
  represent 1.6\% of all falls \citep{2007-AREPS-35-Scott}.
  Together with CI carbonaceous chondrites, they represent the most
  chemically primitive meteorites \citep[i.e., closest to the solar
    composition,][]{2007-AREPS-35-Scott} while paradoxically 
  having suffered extensive hydration 
  \citep[e.g.,][]{2013-GeCoA-123-Alexander}.
  This aqueous alteration took place at low temperature, and
  thermal alteration of CM parent bodies peaked around
  120\degr C
  \citep[e.g., ][]{1962-GeCoA-26-Dufresne,
    1989-Icarus-78-Zolensky, 1997-GeCoA-61-Zolensky,
    2007-GeCoA-71-Guo}, and went up to 
  150\degr C as revealed by the formation of 
  dolomite carbonates \citep{2014-GeCoA-144-Lee}.
  As such, they are thought to have formed from a
  mixture of ice and dust where water ice was subsequently brought to
  a liquid state via the radioactive decay of $^{26}$Al contained in dust
  particles, leading to the aqueous alteration of a significant
  fraction of the dust \citep[see, e.g.,][]{2006-MESS2-Krot}.

  \indent Recently, \citet{2016-AJ-152-Vernazza} made a spectral survey
  of 70 Ch/Cgh asteroids, which included large 200+\,km
  deemed-primordial bodies, but also objects as small
  as 15\,km diameter, presumed to be the collisional fragments
  of dynamical families created from large bodies.  This
  allowed a probe of the internal composition of prior,
  larger parent bodies. 
  \add{These authors interpreted the spectral diversity, already
    reported elsewhere
    \citep[e.g.,][]{1994-Icarus-111-Vilas, 1999-AA-135-Fornasier,
      2014-Icarus-233-Fornasier},
    as resulting mainly from a variation of the average
    regolith grain size rather than to different thermal histories
    \citep{2014-Icarus-233-Fornasier, 2015-AJ-150-Rivkin}, although
    differences in band depths and centers also argue for some
    heterogeneity in mineral abundances
     \citep{1998-MPS-33-Burbine, 2011-Icarus-216-Cloutis,
       2014-Icarus-233-Fornasier}. 
  This evidence points toward an overall homogeneous internal structure for
  the parent bodies of CM chondrites.}

  \indent The thermal modeling of the early internal
  structure of CM parent bodies by \citet{2017-SciAdv-3-Bland}
  supports this conclusion. They showed
  that convection may have prevented strong thermal gradients and
  differentiation of material.
  Starting from the accretion of dust and ices
  beyond the snow line \citep{2018-ApJ-854-Scott}, the original
  non-lithified structure of CM parent bodies
  has allowed large-scale circulation of
  material. This model explains both the limited temperature experienced by
  CM parent bodies, and the small-scale heterogeneity
  \add{\citep[such as temperature and redox state observed at the hundred of
  micrometer level, see][]{2015-GeCoA-161-Fujiya}} observed in CM chondrites
  \citep{2007-GeCoA-71-Guo, 2017-SciAdv-3-Bland}.

  \indent \add{To test this model, we sought to gather evidence for or
    against an originally differentiated internal structure within large
    Ch/Cgh asteroids.
    Our goal was to measure their bulk density, which, when compared
    with the density of CM chondrites, can reveal the presence, or
    absence, of denser material in their interior.
    Determination of a density requires both a mass and
    a volume.  Masses for larger objects can be determined
    by their gravitational effects on other bodies, but
    the precision is often low, and is particularly
    problematic for objects of low mass \citep[see][for a discussion on the precision and
    biases in mass determination]{2012-PSS-73-Carry}.
    If any object is binary, however, the orbit of the
    satellite yields the primary mass directly.  Therefore,
    binary asteroids are crucial to establish  
    solid and accurate references on which a larger population can be
    analyzed.}

  \indent The present article focuses on asteroid
  (41) Daphne as representative of the Ch spectral class,
  around which a satellite \add{was discovered} in 2008
  \citep[\SatOne, which we call
  \NameOne\,\footnote{pronounced ``peh-NEH-oss''},][]{2008-IAUC-8930-Conrad}.
  Daphne was observed within a survey we are currently conducting
  (ID \href{http://archive.eso.org/wdb/wdb/eso/abstract/query?ID=9900740&progid=199.C-0074(A)}{199.C-0074}, PI P. Vernazza)
  to image a substantial fraction of main-belt
  asteroids larger than 100\,km in diameter, sampling the main
  compositional classes
  \citep[see][for a description of
    the survey]{2018-AA-618-Vernazza}.
  We image these asteroids throughout their rotation at high
  angular-resolution with the 
  SPHERE/ZIMPOL extreme adaptive-optics (AO) camera
  \citep{2008-SPIE-Beuzit, 2008-SPIE-Thalmann}
  mounted on the European Southern Observatory (ESO)
  Very Large Telescope (VLT). 
  The high-quality correction delivered by the AO system of SPHERE
  \citep{2006-OExpr-14-Fusco,2014-SPIE-Fusco} compared to previous
  generations of AO cameras, and the use of 
  shorter wavelength (visible R band compared to the near-infrared
  J/H/K bands typically used \add{with} previous cameras), provides a twofold to
  threefold improvement in angular resolution.
  This sharper resolution allows the detailed modeling of 
  asteroid shapes
  \add{and enhanced satellite detection capability, 
    as recently illustrated
    on the
    main-belt asteroids (3) Juno, (6) Hebe,
    (16) Psyche, (89) Julia, (107) Camilla, and (130) Elektra
    \citep{2015-AA-576-Viikinkoski, 2018-AA-Viikinkoski,
      2016-IAUC-Marsset, 2017-AA-604-Marsset, 2017-Messenger-169-Marsset,
      2016-ApJ-820-Yang, 2017-AA-599-Hanus, 
    2018-Icarus-309-Pajuelo, 2018-AA-618-Vernazza}.}
  
  \indent The article is organized as follows:
  we first describe our observations in Section~\ref{sec:obs},
  followed by the determination of the 3-D shape model
  (Section~\ref{sec:phys})
  \add{, the mutual orbit of Daphne and its satellite
  (Section~\ref{sec:dyn}), and 
   the diameter and spectrum of the satellite itself
   (Section~\ref{sec:sat}).} 
  Based on the density of (41) Daphne and a
  compilation of mass and diameter estimates from the literature, we
  then discuss the internal structure of CM parent bodies in  
  Section~\ref{sec:disc}.

\section{Observations\label{sec:obs}}

  \indent Daphne was imaged with ZIMPOL once in May 2017 and three times in August 2018. We also
  compile \numb{24} epochs of 
  high-angular resolution and high-contrast
  images (Table~\ref{tab:ao})
  from large ground-based telescopes equipped with adaptive-optics (AO)
  cameras: NIRC2 at Keck \citep{2004-AppOpt-43-vanDam}
  and NACO at ESO VLT
  \citep{2003-SPIE-4839-Rousset}.

  \indent We process the ZIMPOL data with the ESO pipeline
  \citep[see details in][]{2018-AA-618-Vernazza}.
  We then reduce all other imaging epochs with the
  same suite of IDL routines for consistency.
  The basic data reduction encompasses
  bad pixel mapping and removal by median interpolation,
  sky subtraction, and flat-field correction, following the steps
  described in 
  \citet{2008-AA-478-Carry}.
  We then use \mistral, a myopic
  deconvolution algorithm 
  optimized for celestial targets with sharp boundaries
  \citep{2002-SPIE-Fusco, 2004-JOSAA-21-Mugnier}, 
  to deconvolve the images and enhance their angular resolution for
  shape modeling purposes.
  The results of this approach have already
  been demonstrated elsewhere
  \citep[e.g.,][]{2006-JGR-111-Witasse, 2014-Icarus-236-Drummond}.
  In parallel, the diffused halo of light surrounding Daphne was 
  removed by subtracting concentric 
  annuli \citep[as described thoroughly in][]{2018-Icarus-309-Pajuelo} 
  for satellite detection.

  \indent We complement this data set with an additional epoch of
  Daphne in May 2017 with the SPHERE
  integral-field spectrograph \citep[IFS,][]{2008-SPIE-Claudi}
  to acquire the near-infrared spectrum of its satellite
  \NameOne.
  This epoch was obtained within the
  ``\textsl{Other Science}'' program 
  (ID: \href{http://archive.eso.org/wdb/wdb/eso/abstract/query?ID=9900980}{099.D-0098},
  PI J.-L.~Beuzit) of the 
  guaranteed-time observations (GTO) of the SPHERE consortium. The aim of
  this program is to illustrate SPHERE's
  capabilities in science topics other than its primary goal: direct
  imaging of planets and disks.
  The IFS data were reduced using the SPHERE consortium pipeline, which
  includes bad pixel removal, sky subtraction, flat-field correction, and
  wavelength calibration.

  \begin{figure}[t]
    \centering
    \includegraphics[width=\hsize]{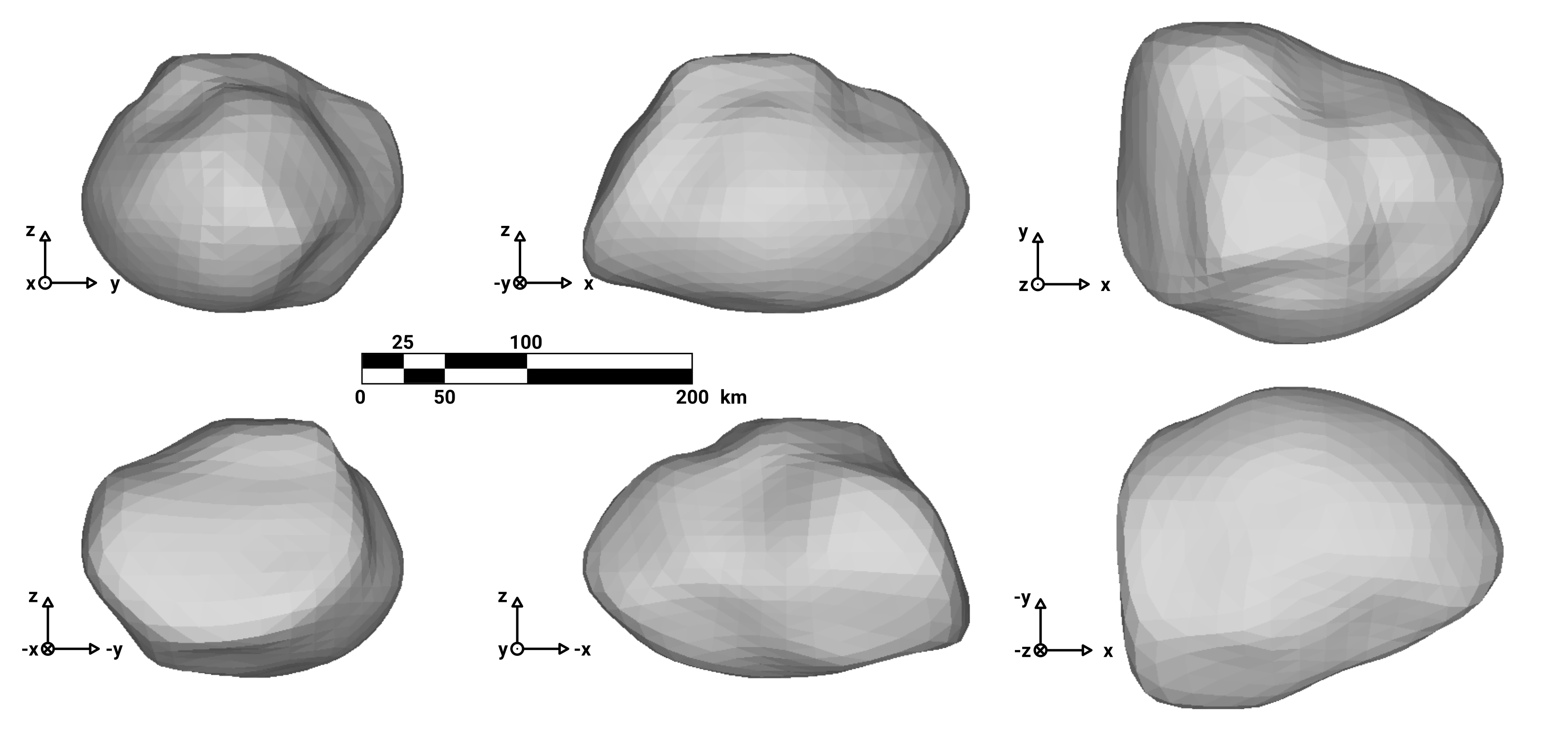}
    \caption[Shape model of Daphne]{%
      Views of the shape model of Daphne.
      The X, Y, and Z axes are aligned
      along the principal axes of inertia. 
    }
    \label{fig:3d}%
  \end{figure}

  \indent Together with these images, we retrieve
  \numb{nine} stellar occultations by Daphne compiled on the PDS by
  \citet{PDSSBN-OCC}.
  We convert the locations of observers and disappearance timings
  into chords on the plane of the sky
  using the recipes by \citet{1999-IMCCE-Berthier}.
  We detail the circumstances of observation of these occultations in
  Table~\ref{tab:occ}. 
  
  \indent We also compile \numb{29} optical lightcurves, from
  the historical works of
  \citet{1977-AA-56-Scaltriti}, 
  \citet{1983-AA-54-Barucci}, 
  \citet{1985-Icarus-61-Barucci}, and 
  \citet{1987-Icarus-70-Weidenschilling, 1990-Icarus-86-Weidenschilling}, used by
  \citet{2002-AA-383-Kaasalainen} and
  \citet{2017-AA-601-Hanus} to reconstruct the 3-D shape of Daphne.
  We complement this data set with 
  \numb{twelve} lightcurves obtained by amateur astronomers,
  \numb{two} lightcurves obtained with the 60\,cm
  \textsl{Andr{\'e} Peyrot} telescope mounted at Les Makes observatory
  on R{\'e}union Island (operated as a partnership among Les Makes
  Observatory and the IMCCE, Paris Observatory),
  \numb{one} lightcurve obtained with the 
  Antarctic Search for Transiting ExoPlanet
  (ASTEP) telescope
  \citep{2010-SPIE-Daban} during its commissioning at the Observatoire
  de la C{\^o}te d'Azur in Nice \citep{2011-Icarus-215-Matter}, 
  and extract \numb{three} serendipitously observed lightcurves from the
  SuperWASP image 
  archive \citep{2017-ACM-Grice}.
  The details of these  lightcurves are provided in Table~\ref{tab:lc}.

\subsection{Properties of Daphne and its satellite}
\subsection{Spin and 3-D shape\label{sec:phys}}

    \indent We determine the spin properties (rotation period and spin-vector
    coordinates) and reconstruct the 3-D shape of Daphne with the 
    open-source\footnote{\url{https://github.com/matvii/adam}}
    \adam~algorithm \citep{2015-AA-576-Viikinkoski}. \adam~uses
    the Levenberg-Marquardt optimization algorithm to find
    the spin and 3-D shape that best reproduce the lightcurves, stellar
    occultation chords, and disk-resolved images simultaneously
    by comparing the observations with synthetic data generated by
    the model at each step. 

\begin{table}[t]
\begin{center}
  \caption{Spin solution \add{(coordinates in ecliptic and
      equatorial J2000 reference frames)} and shape model parameters
    \add{(the overall shape is reported as the $a>b>c$ diameters of a
      triaxial ellipsoid fit to the shape model)}. All uncertainties
    are reported at 3\,$\sigma$.
    \label{tab:shape}
  }
  \begin{tabular}{lllll}
    \hline\hline
    Parameter & Symbol & Value & Unc. & Unit \\
    \hline
    Sidereal period & $P_s$     & 5.987981 & 6.10$^{-5}$ & hour \\
    Longitude       &$\lambda$  & 199.4 & 5. & deg. \\
    Latitude        & $\beta$   & -31.9 & 5. & deg. \\
    Right ascencion & $\alpha$  & 183.5 & 5. & deg. \\
    Declination     & $\delta$  & -36.6 & 5. & deg. \\
    Ref. epoch      & T$_0$     & 2444771.750 & & \\
    \hline
    Diameter   & $\mathcal{D}$   & 187 & 21.5 & km \\
    Volume     & V   & 3.39 $\cdot 10^{6}$ & 7.1 $\cdot 10^{5}$ & km$^3$ \\
    Diam. a    & a   & 237.9 & 21.5 & km\\
    Diam. b    & b   & 184.8 & 21.5 & km\\
    Diam. c    & c   & 156.0 & 21.5 & km\\
    Axes ratio & a/b & 1.29 & 0.19 & \\
    Axes ratio & b/c & 1.18 & 0.21 & \\
    Axes ratio & a/c & 1.52 & 0.25 & \\
    \hline
  \end{tabular}
\end{center}
\end{table}

    \indent The best-fit solution displayed in Fig.~\ref{fig:3d},
    \add{the details of which} are listed in 
    Table~\ref{tab:shape}, has a sidereal rotation period $P_s$ of
    \numb{\sPer}\,h and spin-vector \add{ecliptic} coordinates of
    (\numb{\sLon\degr, \sLat\degr}), very similar to the results from
    convex shape modeling from lightcurves only by 
    \citet{2002-AA-383-Kaasalainen} and
    multi-data shape reconstruction from a slightly different data set
    with \adam~by \citet{2017-AA-601-Hanus}.
    The shape model
    is of an irregular
    body with several large \add{flat areas, putative impact basins,} and 
    has a spherical-volume-equivalent diameter $\mathcal{D}$
    of \numb{\Diam\,$\pm$\,\dDiam}\,km
    (3\,$\sigma$ uncertainty).
    We present all the data compared with the predictions from the
    shape model in Appendix~\ref{app:data}.

    \indent An almost similar shape model and spin solution had been
    previously reported by \citet{2009-PhD-Carry}
    with the multi-data \koala~shape
    reconstruction algorithm
    \citep[\add{based on the same algorithm as \adam{} but with
        different implementation},][]{2010-Icarus-205-Carry-a, 2011-IPI-5-Kaasalainen}.
    Because \koala~has been validated by comparing the 
    shape model of (21) Lutetia 
    \citep{2010-AA-523-Carry, 2010-AA-523-Drummond} with 
    the images returned by the ESA Rosetta mission during its flyby of
    the asteroid \citep{2011-Science-334-Sierks, 2012-PSS-66-Carry},
    the agreement between the two models provides solid evidence for
    the reliability of both \adam~and of the shape model of Daphne. 

    \indent While there is a significant spread of diameter estimates
    in the literature (average diameter of \numb{192\,$\pm$\,38}\,km),
    all the estimates based on direct measurements, i.e. \add{stellar
      occultations, disk-resolved imaging, mid-infrared
      interferometry},
    are narrowly clustered around our value:
    \numb{188\,$\pm$\,14}\,km
    (3\,$\sigma$ deviation, see Table~\ref{tab:diam}).
    In particular, the mid-infrared interferometric observations by
    \citet{2011-Icarus-215-Matter} provide an independent
    confirmation, being based on a totally different data set.
    These authors analyzed their interferometric visibilities using
    the convex shape model of 
    \citet{2002-AA-383-Kaasalainen}
    and the non-convex shape model of 
    \citet[][]{2009-PhD-Carry}, almost identical to the model presented
    here. The associated diameters differ by 15\,km, showing how the
    determination of diameter is sensitive to the shape of the object.
    Their best-fit diameter associated with the non-convex model was
    \numb{185.5\,$\pm$\,10.5}\,km (3\,$\sigma$), supporting the
    present value. 

\begin{figure}[t]
    \centering
    \includegraphics[width=.49\textwidth]{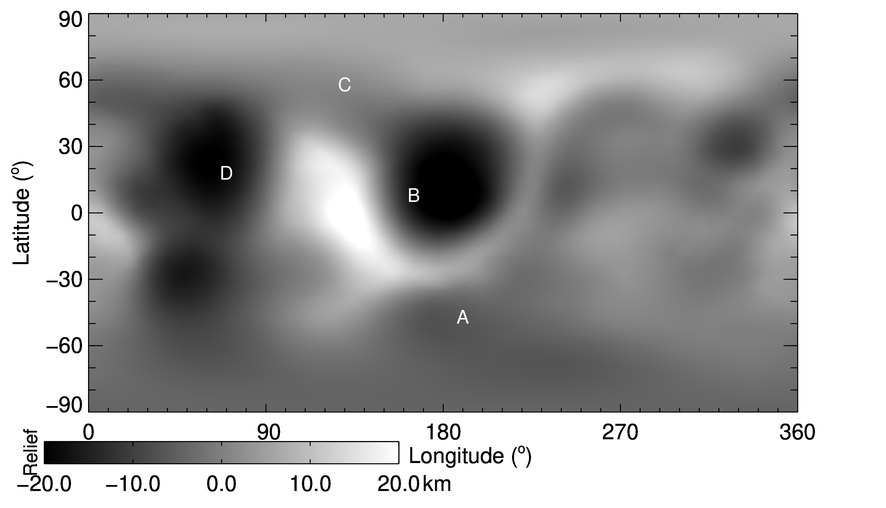}
    \caption[Topography of Daphne]{%
      \add{Topography of Daphne, measured in kilometers with respect to its
        reference ellipsoid (Table~\ref{tab:shape}).
        The putative depression is marked by the letter D, while the
        three flat areas are labeled A, B, and C. They may appear as circular
        depressions here (in particular B) because the topography is
        measured with respect to an ellipsoid.}
    }
    \label{fig:map}%
  \end{figure}

    \indent \add{We define the prime meridian of Daphne to be along its longest
      axis, on which a hill, which we call the nose, is present at the equator
      (elevation of \numb{11}\,km above the reference ellipsoid, see
      the map in Fig.~\ref{fig:map}).
      Several flat regions (hereafter A, B, C)
      can be identified in the images and on the
      shape model, as well as a clear depression (D).
      The three flat areas are located near both poles, and on the opposite
      side of the nose.
      These flat areas and the depression are large compared with
      the diameter of Daphne and may be indicative of large
      impact basins not modeled because concavities could not be
      detected with the 
      available data \citep{2003-AA-404-Durech,2015-MNRAS-453-Devogele}.
      The central location and overall dimensions of these features, including
      the ratio of their surface-equivalent diameter to the diameter of Daphne,
      are listed in Table~\ref{tab:feat}.
    }

\begin{table}
\begin{center}
  \caption[Notable topographic features]{%
    Dimensions (semi-axes $\mathcal{L}_a$ and $\mathcal{L}_b$,
    surface area $\mathcal{A}$, and fraction $f$ of Daphne's diameter) and planetocentric coordinates
    ($\lambda_c$, $\beta_c$)
    of the four notable topographic features
    of Daphne.}
  \label{tab:feat} 
  \begin{tabular}{l r@{\,$\pm$\,}l r@{\,$\pm$\,}l r@{\,$\pm$\,}l r@{\,$\pm$\,}l  l}
    \hline
    & \multicolumn{2}{c}{A} &
    \multicolumn{2}{c}{B} &
    \multicolumn{2}{c}{C} &
    \multicolumn{2}{c}{D} & Unit\\
    \hline
    $\lambda_c$      & 190 & 5  & 165 & 5 & 130 & 5 &  60 & 5 & deg. \\
    $\beta_c$        & -50 & 5  &  +5 & 5 & +55 & 5 & +19 & 5 & deg. \\
    $\mathcal{L}_a$  &  60 & 5  &  68 & 5 &  45 & 5 &  42 & 3 & km   \\
    $\mathcal{L}_b$  &  48 & 5  &  42 & 5 &  45 & 5 &  42 & 3 & km   \\
    $\mathcal{A}  $  & 8.7 & 3.3& 9.0 &3.3& 6.3 &2.8& 5.5 &0.8& 10$^3$\,km$^2$ \\
    $f$              &  56 & 10 &  57 & 10&  47 & 10&  44 & 3 & \% \\
    \hline
  \end{tabular}
\end{center}
\end{table}

\subsection{Dynamics of the system\label{sec:dyn}}

\begin{figure*}[t]
    \centering
    \includegraphics[width=.8\textwidth]{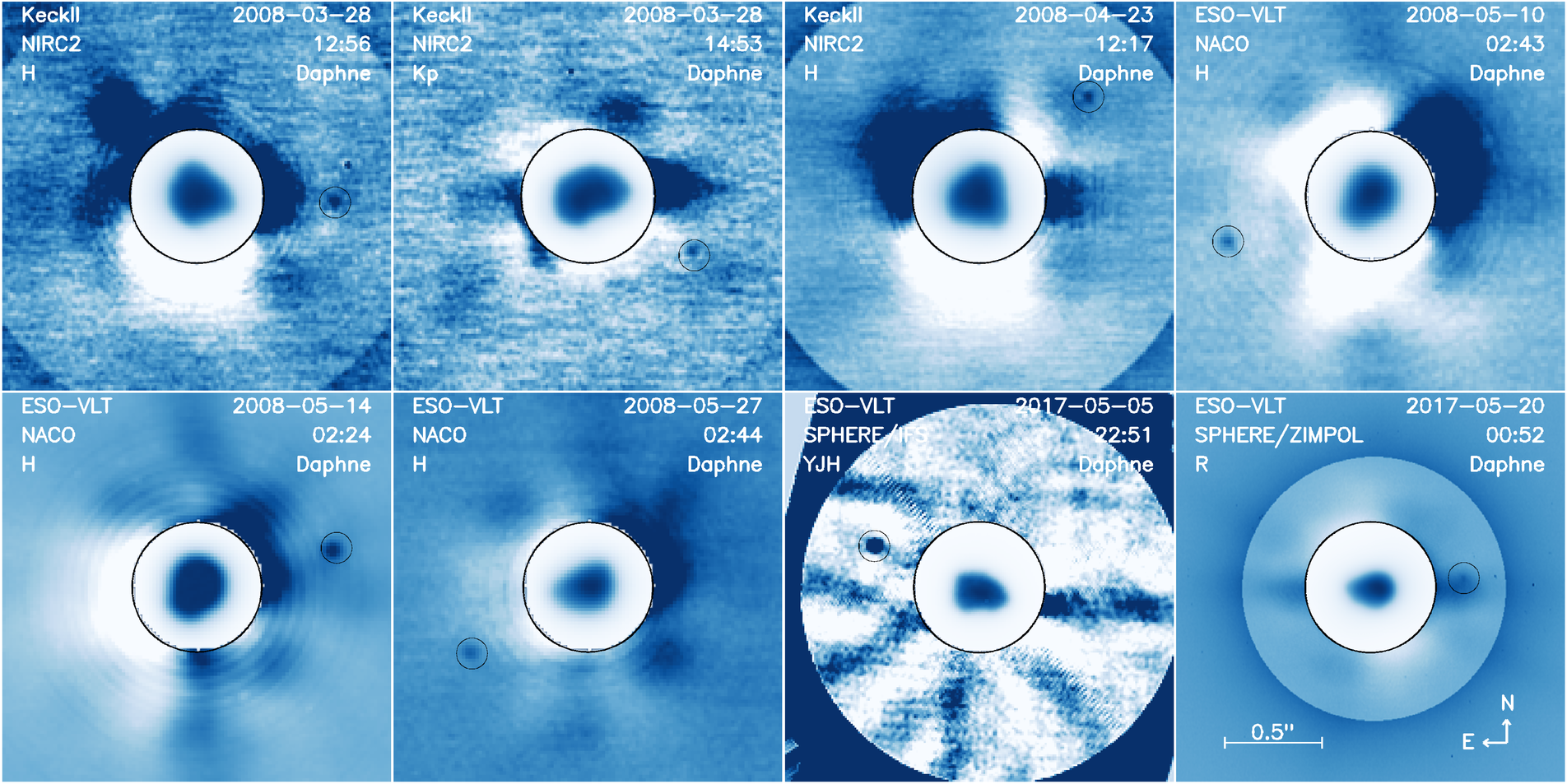}
    \caption[AO images of Daphne]{%
      Example images of Daphne and its satellite.
      Each panel represents a different epoch.
      The image is displayed in the inner circle, showing the irregular shape of Daphne.
      The outer region shows the images after halo removal, showing the satellite (highlighted by a small circle).
      On each panel the telescope, instrument, filter, and UTC date are
      reported.
    }
    \label{fig:bin:img}%
  \end{figure*}

  \indent On each image, we measure the relative position of the
  satellite with respect to center 
  of light of the primary by adjusting a 2-D gaussian on the primary
  using the unmodified images and another on 
  the satellite using the halo-removed images
  (Fig.~\ref{fig:bin:img}).
  We use the meta-heuristic algorithm
  \genoid~\citep{2012-AA-543-Vachier} to find the set of orbital 
  parameters that best fit the observations.
  \add{The orbital  parameter space is usually
    6-D for a simple Keplerian motion: orbital
    period, excentricity, inclination, 
    longitude of the ascending node, 
    argument of periapsis, and time of passage to the periapsis.
    More dimensions are required if the
    gravitational potential is not central, such as
    in accounting for a gravitational quadrupole J$_2$.
    \genoid~explores this parameter space
    in successive generations of orbital solutions},
  randomly merging
  the parameters of the best solutions to create newer generations.
  In combination with this broad exploration of the parameter space,
  \genoid~uses gradient descent 
  at each generation to find the minimum closest to each best-trial solution.

  \indent The reliability of this approach has been assessed during a
  stellar occultation by (87) Sylvia in 2013.
  We had used \genoid~ to predict the position of its largest satellite
  Romulus before the event, placing observers on the occultation path
  of the satellite. Four different observers detected an occultation
  by Romulus at only 13.5\,km off the predicted position
  \citep[][]{2014-Icarus-239-Berthier}. 

  \indent The best-fit orbit adjusts the \numb{\ObsOne}~positions with
  a root-mean square (RMS) residual of \numb{\rmsOne}~mas only, i.e.,
  smaller than the pixel size of most observations
  (\numb{21 out of \ObsOne} taken with NACO/VLT and NIRC2/Keck, see
  Table~\ref{tab:genoid}). The observations, covering 
  \numb{\obsDay}~days
  or 
  \numb{\obsRev}~revolutions, provide solid constraints on the orbital
  period (\numb{\orbPer}\,day).
  The satellite orbits Daphne on a \add{Keplerian, equatorial,} and prograde
  orbit, slightly eccentric (Table~\ref{tab:dyn}), at a distance of
  \add{only \numb{7.4} primary radii.
    We searched for a signature of the gravitational quadrupole
    J$_2$ but the poor time coverage of the astrometry with a gap of
    nine years between the positions in 2008 and the two in 2017
    precluded any firm conclusion. A regular follow-up of the position
    of \SatOne{} is required to conclude on the J$_2$ of Daphne.}
  These \add{orbital} characteristics
  argue in favor of a formation of the
  satellite by impact excavation, and re-accumulation of material in
  orbit, followed by tidal circularization
  \citep{1989-AsteroidsII-III-Weidenschilling,
    2002-AsteroidsIII-2.2-Merline, 2004-Icarus-170-Durda,
    2015-AsteroidsIV-Margot}.

\begin{table}
\begin{center}
  \caption[Orbital elements of \NameOne, the satellite of Daphne]{%
    Orbital
    elements of \NameOne, the satellite of Daphne, 
    expressed in EQJ2000,
    obtained with \genoid:
    orbital period $P$, semi-major axis $a$,
    eccentricity $e$, inclination $i$,
    longitude of the ascending node $\Omega$,
    argument of pericenter $\omega$, time of pericenter $t_p$.
    The number of observations and RMS between predicted and
    observed positions are also provided.
    Finally, we report the derived primary mass $M$,
    the ecliptic J2000 coordinates of the orbital pole
    ($\lambda_p,\,\beta_p$), 
    the equatorial J2000 coordinates of the orbital pole
    ($\alpha_p,\,\delta_p$), and the
    orbital inclination ($\Lambda$) with respect to the equator of
    Daphne. Uncertainties are given at 3\,$\sigma$.}
  \label{tab:dyn} 
  \begin{tabular}{l ll}
    \hline
    \noalign{\smallskip}
    \multicolumn{2}{c}{Observing data set} \\
    \noalign{\smallskip}
    Number of observations & \ObsOne   &  \\
    Time span (days)       & 3783      &  \\
    RMS (mas)              & \rmsOne   &  \\
    \hline
    \noalign{\smallskip}
    \multicolumn{2}{c}{Orbital elements EQJ2000} \\
    \noalign{\smallskip}
    $P$ (day)        &    1.137\,446 & $\pm$ 0.000\,009 \\
    $a$ (km)         &      463.5    & $\pm$ 22.8      \\
    $e$              &      0.009    & $_{-0.009}^{+0.021}$  \\
    $i$ (\degr)      &     128.1     & $\pm$  6.2      \\
    $\Omega$ (\degr) &     272.7     & $\pm$  5.7      \\
    $\omega$ (\degr) &     171.9     & $\pm$  37.6     \\
    $t_{p}$ (JD)     & 2454550.586985& $\pm$ 0.117331   \\
    \hline
    \noalign{\smallskip}
    \multicolumn{2}{c}{Derived parameters} \\
    \noalign{\smallskip}
    $M$ ($\times 10^{18}$ kg)     &  6.10   & $\pm$ 0.89    \\
    $\lambda_p,\,\beta_p$ (\degr) &197, -33   & $\pm$ 6, 7 \\
    $\alpha_p,\,\delta_p$ (\degr) &182, -38   & $\pm$ 5, 7 \\
    $\Lambda$ (\degr)             &  2        & $\pm$ 4    \\
    \hline
  \end{tabular}
\end{center}
\end{table}

\subsection{Spectrum and diameter of the satellite\label{sec:sat}}


  \indent \add{We recorded the near-infrared spectra of Daphne and its
    satellite with SPHERE/IFS. 
    Telluric features were removed by observing the nearby star
    HD 102085 (G3V). 
    Similarly to previous sections, the bright halo of
    Daphne that contaminated the spectrum of the moon was
    removed. This was 
    achieved by measuring the background at the location of the
    moon for each pixel as the median value of the area defined as
    a 40\,$\times$1-pixel arc centered on Daphne.
    To estimate the
    uncertainty and potential bias on photometry (at each wavelength)
    introduced by this
    method, we performed a number of simulations \citep[similarly to our
    study of the satellite of Camilla,
    see][]{2018-Icarus-309-Pajuelo}.}
  
  \indent \add{
    We added synthetic companions on the 39 spectral images
    of the spectro-imaging cube, at a similar separation to the
    satellite ($\approx$\numb{300}\,mas) and
    random position angles from the primary. The simulated sources
    were modeled as the point-spread function from the calibration
    star images scaled in brightness. 
    The halo from Daphne was then removed from these simulated
    images using the method described above, and the flux of the
    simulated companion measured by adjusting a 2D-Gaussian
    profile (see Sect.~\ref{sec:dyn}).
    Based on a total statistics of \numb{100} simulated
    companions, we find that the median loss of flux at each
    wavelength is 13\,$\pm$\,9\% and that the spectral slope is 
    affected (\numb{-0.50\,\%/100nm} on average).} 

  \indent \add{The measured slope for each individual simulation,
    however, is not reliable, the standard deviation of all simulations
    being of \numb{2.73}\,\%/100nm. The value for each depends on the
    location (position angle) of the simulated satellite. 
    Furthermore, the signal-to-noise ratio of the simulations range
    from virtually zero to four, with an average of 1.8 only.
    The spectrum of the satellite, therefore, is not reliable, being
    noisy and likely affected by strong slope effects. }
  
  \indent \add{From the integrated fluxes measured with the 2-D
    gaussian functions on Daphne and its satellite at each epoch
    (Fig.~\ref{fig:dmag} and Table~\ref{tab:genoid}), we
    derived their magnitude difference to be
    \numb{\DMag}.
    Combined with our determination of a diameter of
    \numb{\Diam\,$\pm$\,\dDiam}\,km, 
    we derive a diameter of}
  \numb{\satDiam}\,km for the satellite, under the
  assumption of a similar albedo for both.
  This assumption is supported by
  the multiple reports of spectral similarities
  between 
  the components of multiple asteroid systems:
  (22) Kalliope \citep{2009-Icarus-204-Laver},
  (90) Antiope \citep{2009-MPS-44-Polishook},
  (107) Camilla \citep{2018-Icarus-309-Pajuelo},
  (130) Elektra \citep{2016-ApJ-820-Yang}, and
  (379) Huenna \citep{2011-Icarus-212-DeMeo}.

\section{Implications on the internal structure\label{sec:disc}}

  \indent \add{Using the volume determined from
    \adam~(Sect.~\ref{sec:phys}) and
    the mass from \genoid~(Sect.~\ref{sec:dyn}), we
    derive a bulk density 
    of \numb{\Dens\,$\pm$\dDens} (\sid, 3\,$\sigma$ uncertainty).
    This value} is smaller but marginally consistent with the typical
  density of  
  the CM carbonaceous chondrites, at
  \numb{2.13\,$\pm$0.57}\,\sid{} 
  \citep[3\,$\sigma$,][]{2008-ChEG-68-Consolmagno}.
  \add{Comparing these two density values 
    implies a macroporosity of \numb{\Poro}\,$\pm$\,\dPoro\%.}

\begin{figure*}[t]
    \centering
    \includegraphics[width=\hsize]{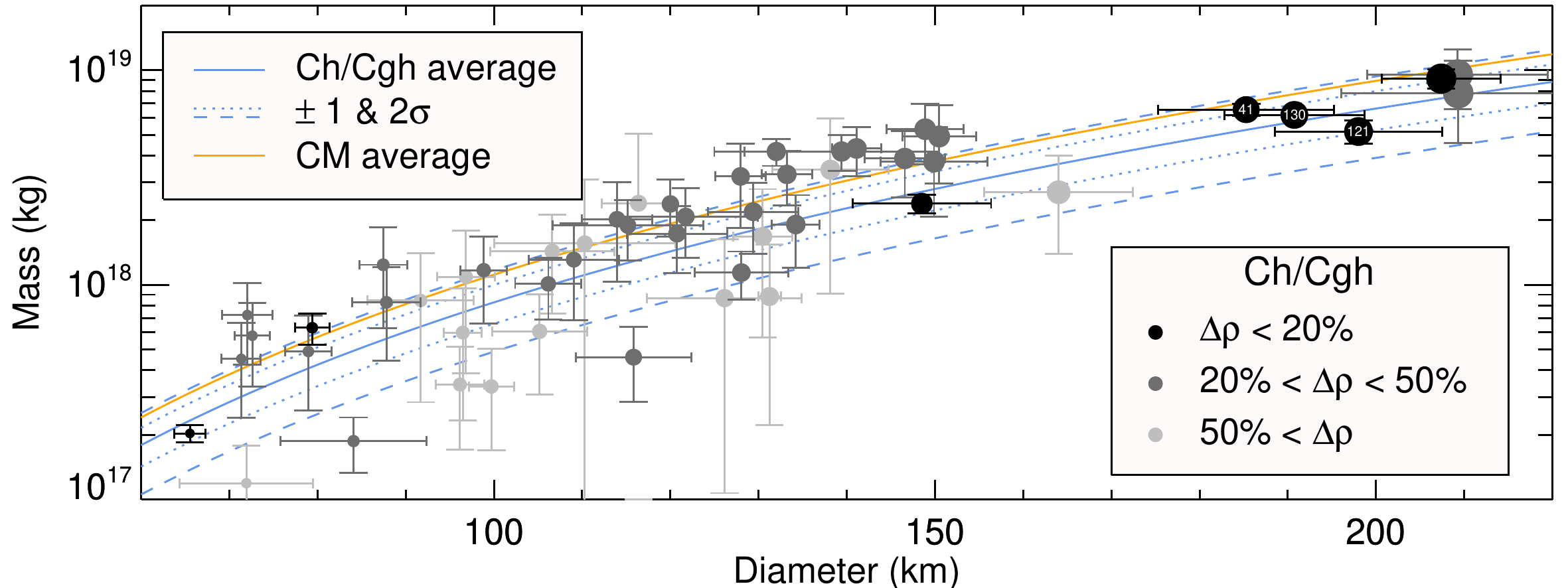}
    \caption[Mass-Diameter for Ch/Cgh asteroids]{%
      Mass vs diameter distribution for \numb{76} Ch/Cgh asteroids
      \add{(grey circles).
        Those more precise than
        50\% and 20\% are plotted as dark grey and black circles,
        respectively.
        Unreliable density estimates (below 0.5\,\sid{}) are not
        displayed.}
      The three binary asteroids,
      (41) Daphne, (121) Hermione, (130) Elektra, are labeled.
      The solid and dashed blue curves represent the average density
      (\numb{1.58}\,\sid) and
      1\,$\sigma$--2\,$\sigma$
      (\numb{$\pm$\,0.33}\,\sid) dispersion of the sample.
      The solid orange curve stands for the average density of CM
      chondrites at \numb{2.13\,$\pm$\,0.57}\,\sid{} (3\,$\sigma$).
    }
    \label{fig:cgh}%
  \end{figure*}

  As visible in Fig.~\ref{fig:cgh}, Ch
  and Cgh asteroids over two orders of magnitude in mass follow the
  same trend and have a density smaller than that of CM meteorites, 
  although there is definitively some scatter due to 
  large uncertainties and biases in the determination of their
  density 
  \citep[see Appendix~\ref{app:cgh} for the complete list of estimates,
    and][for a discussion on the reliability of mass, diameter, and
    density estimates]{2012-PSS-73-Carry}.
  The distribution of all density estimates weighted by their
  respective uncertainty is Gaussian with an average value
  of
  \numb{1.40$_{-1.40}^{+1.92}$}\,\sid{}.
  This value changes to \numb{1.58\,$\pm$\,0.97}\,\sid{} 
  (3\,$\sigma$) by
  considering only the estimates with less than 20\% relative
  uncertainty
  (represented by the blue curves in Fig.~\ref{fig:cgh}).
  This highlights the importance of the study of binary systems that
  can be angularly resolved to accurately determine both the volume
  and the mass, hence the density.

  \indent There are three large (150+\,km) binary Ch-type asteroids:
  (41) Daphne, (121) Hermione, and (130) Elektra.
  The density of Hermione was estimated to
  \numb{1.4$^{+1.5}_{-0.6}$}\,\sid{} by
  \citet{2009-Icarus-203-Descamps}
  and 
  \numb{1.26\,$\pm$\,0.90}\,\sid{} by
  \citet{2017-AA-607-Viikinkoski}.
  The density of Elektra was reported as
  \numb{1.60\,$\pm$\,0.39}\,\sid{} by \citet{2017-AA-599-Hanus}, based
  on the mass estimate by \citet{2008-Icarus-196-Marchis} while
  the mass recently derived by
  \citet{2016-ApJ-820-Yang} leads to \numb{1.50\,$\pm$\,0.36}\,\sid{}.
  These values are somewhat smaller than the density of the three other large
  Ch-types:
  (13) Egeria at \numb{1.99\,$\pm$\,0.69}\,\sid{}, 
  (19) Fortuna at \numb{1.96\,$\pm$\,0.28}\,\sid{}, and
  (48) Doris at \numb{1.63\,$\pm$\,0.74}\,\sid{}
  (see Appendix~\ref{app:cgh}).
  The density of these six targets is \add{systematically smaller
  than CM meteorites}. 

  \indent Over the wide range of mass spanned by the Ch/Cgh asteroids
  studied here (Fig.~\ref{fig:cgh}), the density appears roughly
  constant although there is a larger spread toward smaller diameters.
  This larger spread is likely due to larger uncertainties on the density
  estimates \citep[the gravitational signature of asteroids become
    harder to detect toward smaller sizes, see][]{2012-PSS-73-Carry}.
  Indeed, the report by \citet{2012-PSS-73-Carry}
  of a correlation 
  between density and mass (or diameter)
  due to an
  increasing macroporosity toward smaller diameters mainly happens for
  objects smaller than 100\,km.
  This density of Ch/Cgh asteroids systematically 
  lower than that of CM meteorites has two
  implications.

  \indent First, the density of the largest Ch/Cgh asteroids argues
  against the presence of material denser than CM-like material in
  their interior, i.e., differentiation.
  \add{If denser material was present, their macroporosity would have
    to be large to maintain their bulk density, contrarily to what is
    observed for 100-200+\,km bodies \citep{2012-PSS-73-Carry}.}
  Together with the observed compositional homogeneity of 
  Ch/Cgh asteroids \citep{2016-AJ-152-Vernazza}, 
  it supports the idea
  of an originally
  \rem{homogeneous}\add{undeifferentiated} internal structure of CM parent bodies.
  Because collisional fragments sample the interior of their parent
  bodies, this spectral homogeneity only modulated by grain sizes
  together with the density smaller than the surface analog material
  can be  
  interpreted as evidence for a homogeneous internal
  structure, without differentiation, of the CM parent bodies.
  The thermal modeling in ``giant mud balls'' (non-lithified structure)
  proposed by \citet{2017-SciAdv-3-Bland}  then 
  provides an explanation for the observed peak temperature of 120\degr C
  \citep[e.g., ][]{2007-GeCoA-71-Guo}.

  \indent Second,
  the density of Ch/Cgh asteroids being systematically smaller than CM
  chondrite indicates the presence of large voids, or less dense
  material, in their interior.
  Given the extensive hydration suffered by CM chondrites
  \citep[e.g.,][]{2013-GeCoA-123-Alexander} and the detection of
  water ice on/near the surface of asteroids well-within the snowline
  \citep[e.g.,][]{2010-Nature-464-Campins, 2014-Nature-505-Kueppers}, 
  one cannot reject the hypothesis of the presence of water ice in
  Ch/Cgh asteroids.
  However, considering that all these objects have experienced
  numerous collisions over the history of the solar system, some
  macroporosity, i.e. voids, can be expected.


\section{Conclusion\label{sec:conclu}}

  \indent We have acquired high-angular resolution images and
  spectro-images \add{of the binary
  Ch-type \add{asteroid} (41) Daphne} from 
  large ground-based telescopes, including the new generation extreme
  adaptive-optics SPHERE camera mounted on the ESO VLT.
  We determined the orbit of its small satellite \NameOne{} from \numb{\ObsOne{}}
  observations of its position around Daphne, with an RMS of 
  \rmsOne~mas only. 
  Combining our disk-resolved images with optical lightcurves and
  stellar occultations, we determined the 3-D shape of 
  (41) Daphne. 

  \indent The derived density of
  \numb{\Dens\,$\pm$\dDens}\,\sid{} provides a solid reference to
  interpret the density of Ch/Cgh asteroids.
  The density is available for \numb{76} of them and
  is found systematically smaller than that of the
  associated CM carbonaceous chondrites
  \add{(2.13\,$\pm$\,0.57\,\sid{}, 3\,$\sigma$)}, including for the largest
  asteroids.
  This provides robust evidence for a primordial homogeneous internal structure
  of these asteroids, in agreement with the observation of
  compositionnal homogeneity among Ch/Cgh 
  asteroids of very different diameters by
  \citet{2016-AJ-152-Vernazza} and the modeling of their early thermal
  history by \citet{2017-SciAdv-3-Bland}.

\section*{Acknowledgements}

  \indent Some of the work presented here is
  based on observations collected at the European Organisation
  for Astronomical Research in the Southern Hemisphere under ESO
  programs
  \href{http://archive.eso.org/wdb/wdb/eso/abstract/query?ID=8150110}{281.C-5011}
  (PI Dumas),
  \href{http://archive.eso.org/wdb/wdb/eso/abstract/query?ID=9900980}{099.D-0098},
  (SPHERE GTO),
  and
  \href{http://archive.eso.org/wdb/wdb/eso/abstract/query?ID=9900740&progid=199.C-0074(A)}{199.C-0074(A)}
  (PI Vernazza).

  \indent Some of the data presented herein were obtained at the W.M.
  Keck Observatory, which is operated as a scientific partnership
  among the California Institute of Technology, the University of
  California and the National Aeronautics and Space
  Administration. The 
  Observatory was made possible by the generous financial support
  of the W.M. Keck Foundation.

  \indent  This research has made use of the Keck Observatory Archive
  (KOA), which is operated by the W. M. Keck Observatory and the
  NASA Exoplanet Science Institute (NExScI), under contract with the
  National Aeronautics and Space Administration.

  \indent \add{We thank the AGORA association which administrates the
    60 cm telescope at Les Makes observatory, under a financial
    agreement with Paris Observatory. Thanks to A. Peyrot, J.-P. Teng
    for local support, and A. Klotz for helping with the robotizing.} 

  \indent Some of these observations were acquired under grants from
  the National Science Foundation and NASA to Merline (PI).
  B. Carry, A. Drouard, J. Grice and P. Vernazza were supported by
  CNRS/INSU/PNP. 
  J. Hanus and J. Durech were supported by the grant 18-09470S of the
  Czech Science Foundation. 
  The research leading to these results has received funding from the European Union's
  Horizon 2020 Research and Innovation Programme, under Grant
  Agreement no 687378.

  \indent This paper makes use of data from the DR1 of the WASP data
  \citep{2010-AA-520-Butters} as provided by the WASP consortium, 
  and the computing and storage facilities at the CERIT Scientific
  Cloud, reg. no. CZ.1.05/3.2.00/08.0144 
  which is operated by Masaryk University, Czech Republic. 

  \indent \add{TRAPPIST-South is funded by the Belgian Fund for Scientific Research
    (Fond National de la Recherche Scientifique, FNRS) under the grant FRFC
    2.5.594.09.F, with the participation of the Swiss FNS. TRAPPIST-North is a
    project funded by the University of Liège, and performed in collaboration with
    Cadi Ayyad University of Marrakesh. E. Jehin and M. Guillon are Belgian FNRS Senior Research
    Associates.}

  \indent The authors wish to recognize and acknowledge the very sig-
  nificant cultural role and reverence that the summit of Mauna Kea
  has always had within the indigenous Hawaiian community. We
  are most fortunate to have the opportunity to conduct observations
  from this mountain.
  
  \indent Thanks to all the amateurs worldwide who regularly observe
  asteroid lightcurves and stellar occultations. Some co-authors of
  this study are amateurs who observed Daphne, and provided crucial data.

  \indent The authors acknowledge the use of the Virtual Observatory
  tools
  Miriade\,\footnote{Miriade: \href{http://vo.imcce.fr/webservices/miriade/}{http://vo.imcce.fr/webservices/miriade/}}
  \citep{2008-ACM-Berthier}, 
  TOPCAT\,\footnote{TOPCAT:
    \href{http://www.star.bris.ac.uk/~mbt/topcat/}{http://www.star.bris.ac.uk/~mbt/topcat/}}, and
  STILTS\,\footnote{STILTS: \href{http://www.star.bris.ac.uk/~mbt/stilts/}{http://www.star.bris.ac.uk/~mbt/stilts/}}
  \citep{2005-ASPC-Taylor}. This research used the
  SSOIS\,\footnote{SSOIS:
    \href{http://www.cadc-ccda.hia-iha.nrc-cnrc.gc.ca/en/ssois}{http://www.cadc-ccda.hia-iha.nrc-cnrc.gc.ca/en/ssois}}
  facility of the Canadian Astronomy Data Centre operated by the
  National Research Council of Canada with the support of the Canadian
  Space Agency \citep{2012-PASP-124-Gwyn}.

  \indent Part of the data utilized in this publication were
  obtained and made available by the MIT-UH-IRTF Joint Campaign
  for NEO Reconnaissance, using SpeX spectrograph
  \citep{2003-PASP-115-Rayner}. The IRTF is operated by the University of
  Hawaii under Cooperative Agreement no. NCC 5-538 with the National
  Aeronautics and Space Administration, Office of Space Science,
  Planetary Astronomy Program. The MIT component of this work is
  supported by NASA grant 09-NEOO009-0001, and by the National Science
  Foundation under Grants Nos. 0506716 and 0907766. 

  \indent \add{We wish to acknowledge Professor Robert Groves,
    Director of Basic Languages (Classics) of the
    University of Arizona, for his assistance in selecting the name for
    Daphne's satellite. }

  \bibliographystyle{aa} 
  \bibliography{current}

\clearpage
\appendix

\section{Compilation of diameter and mass estimates\label{app:ssodd}}

  \begin{table}[h]
\begin{center}
  \caption[Diameter estimates of (41) Daphne]{
    The diameter estimates ($\mathcal{D}$) of (41) Daphne collected in the literature.
    For each, the 3\,$\sigma$ uncertainty, method, and 
    bibliographic reference are reported. The methods are
    \textsc{adam}: Multidata 3-D Modeling, \textsc{lcimg}: 3-D Model scaled with Imaging, \textsc{lcocc}: 3-D Model scaled with Occultation, \textsc{neatm}: Near-Earth Asteroid Thermal Model, \textsc{stm}: Standard Thermal Model, \textsc{tpm}: Thermophysical Model.
    \label{tab:diam}
  }
  \begin{tabular}{rrrll}
    \hline\hline
     \multicolumn{1}{c}{\#} & \multicolumn{1}{c}{$\mathcal{D}$} & \multicolumn{1}{c}{$\delta \mathcal{D}$} &
     \multicolumn{1}{c}{Method} & \multicolumn{1}{c}{Reference}  \\
    & \multicolumn{1}{c}{(km)} & \multicolumn{1}{c}{(km)} \\
    \hline
  1 &    203.00 &   60.90 & \textsc{stm}   & \citet{PDSSBN-TRIAD}                     \\
  2 &    174.00 &   35.10 & \textsc{stm}   & \citet{PDSSBN-IRAS}                      \\
  3 &    172.43 &   12.24 & \textsc{stm}   & \citet{2010-AJ-140-Ryan}                 \\
  4 &    207.87 &   31.56 & \textsc{neatm} & \citet{2010-AJ-140-Ryan}                 \\
  5 &    187.00 &   60.00 & \textsc{lcocc} & \citet{2011-Icarus-214-Durech}           \\
  6 &    201.50 &   22.50 & \textsc{tpm}   & \citet{2011-Icarus-215-Matter}           \\
  7 &    185.50 &   10.50 & \textsc{tpm}   & \citet{2011-Icarus-215-Matter}           \\
  8 &    179.61 &    7.74 & \textsc{stm}   & \citet{2011-PASJ-63-Usui}                \\
  9 &    205.50 &    5.64 & \textsc{neatm} & \citet{2012-ApJ-759-Masiero}             \\
 10 &    186.00 &   81.00 & \textsc{lcimg} & \citet{2013-Icarus-226-Hanus}            \\
 11 &    198.74 &  185.13 & \textsc{neatm} & \citet{2015-ApJ-814-Nugent}              \\
 12 &    188.00 &   15.00 & \textsc{adam}  & \citet{2017-AA-601-Hanus}                \\
 13 &    187.00 &   21.50 & \textsc{adam}  & This work                        \\
\hline
  \end{tabular}
\end{center}
\end{table}

  \begin{table}[h]
\begin{center}
  \caption[Mass estimates of (41) Daphne]{
    The mass estimates ($\mathcal{M}$) of (41) Daphne collected in the literature.
    For each, the 3\,$\sigma$ uncertainty, method, and 
    bibliographic reference are reported. The methods are
    \textsc{bgeno}: Binary: Genoid, \textsc{defl}: Deflection, \textsc{ephem}: Ephemeris.
    \label{tab:mass}
  }
  \begin{tabular}{rrll}
    \hline\hline
     \multicolumn{1}{c}{\#} & \multicolumn{1}{c}{Mass ($\mathcal{M}$)} &
     \multicolumn{1}{c}{Method} & \multicolumn{1}{c}{Reference}  \\
    & \multicolumn{1}{c}{($\times 10^{18}$ kg)} \\
    \hline
  1 & $10.50 \pm 2.99$               & \textsc{ephem}& \citet{2009-AA-507-Fienga}               \\
  2 & $7.90 \pm 2.37$                & \textsc{ephem}& \citet{2009-SciNote-Folkner}             \\
  3 & $8.43_{-8.43}^{+10.56}$        & \textsc{ephem} & \citet{2011-Icarus-211-Konopliv}         \\
  4 & $18.20_{-18.20}^{+21.60}$      & \textsc{defl}   & \citet{2011-AJ-142-Zielenbach}           \\
  5 & $0.30_{-0.30}^{+17.01}$        & \textsc{defl}   & \citet{2011-AJ-142-Zielenbach}           \\
  6 & $4.76_{-4.76}^{+16.50}$        & \textsc{defl}   & \citet{2011-AJ-142-Zielenbach}           \\
  7 & $12.10_{-12.10}^{+31.50}$      & \textsc{defl}   & \citet{2011-AJ-142-Zielenbach}           \\
  8 & $10.20 \pm 3.57$               & \textsc{ephem}& \citet{2011-DPS-Fienga}                  \\
  9 & $7.79 \pm 5.40$                & \textsc{ephem}& \citet{2013-Icarus-222-Kuchynka}         \\
 10 & $8.29 \pm 2.62$                & \textsc{ephem}& \citet{2013-SoSyR-47-Pitjeva}            \\
 11 & $7.13 \pm 2.01$                & \textsc{ephem}& \citet{2014-SciNote-Fienga}              \\
 12 & $9.35 \pm 4.17$                & \textsc{defl} & \citet{2014-AA-565-Goffin}               \\
 13 & $9.78_{-9.78}^{+13.77}$        & \textsc{defl}   & \citet{2014-SoSyR-48-Kochetova}          \\
 14 & $4.44 \pm 2.52$                & \textsc{ephem}& \citet{2018-INPOP-Fienga}                \\
 15 & $6.10 \pm 0.89$                & \textsc{bgeno}& This work                        \\
\hline
  \end{tabular}
\end{center}
\end{table}

%
%


\clearpage
\section{Shape modeling: data and model predictions\label{app:data}}

\begin{table*}
\begin{center}
  \caption{Date, telescope, camera, number of epochs
    ($\mathcal{N}_S$),
    heliocentric distance ($\Delta$),
    range to observer ($r$),
    phase angle ($\alpha$),
    angular diameter ($\Theta$), and
    \add{program ID and Principal Investigator (PI)}
    for each night of AO imaging observations.
    \label{tab:ao}
  }
  \begin{tabular}{rlllrrrrrll}
    \hline\hline
     & Date & Telescope & Instrument & 
     \multicolumn{1}{c}{$\mathcal{N}_S$} &
     \multicolumn{1}{c}{$\Delta$} &
     \multicolumn{1}{c}{$r$} &
     \multicolumn{1}{c}{$\alpha$} &
     \multicolumn{1}{c}{$\Theta$} & \add{Prog. ID} & PI \\
    &&&&& 
     \multicolumn{1}{c}{(au)} & \multicolumn{1}{c}{(au)} &
     \multicolumn{1}{c}{(\degr)} & \multicolumn{1}{c}{(mas)} \\
    \hline
 1 & 2002-12-29 & Keck  & NIRC2         &  1 & 2.65 & 1.91 & 16.4 & 135 & C74N2 & J.-L. Margot   \\
 2 & 2003-05-06 & Keck  & NIRC2         &  1 & 2.28 & 2.06 & 26.2 & 125 & N17N2 & W. J. Merline  \\
 3 & 2008-01-21 & Keck  & NIRC2         &  7 & 2.19 & 1.78 & 26.3 & 144 & DDT   & A. Conrad \\
 4 & 2008-03-28 & Keck  & NIRC2         & 10 & 2.06 & 1.09 &  9.1 & 235 & DDT   & A. Conrad \\
 5 & 2008-04-23 & Keck  & NIRC2         &  5 & 2.03 & 1.06 &  9.9 & 243 & DDT   & A. Conrad \\
 6 & 2008-05-10 & VLT   & NACO          &  4 & 2.02 & 1.12 & 17.5 & 229 & 281.C-5011 & C. Dumas       \\
 7 & 2008-05-14 & VLT   & NACO          &  2 & 2.02 & 1.14 & 19.1 & 225 & 281.C-5011 & C. Dumas       \\
 8 & 2008-05-27 & VLT   & NACO          &  2 & 2.01 & 1.23 & 23.6 & 209 & 281.C-5011 & C. Dumas       \\
 9 & 2010-11-30 & Keck  & NIRC2         &  1 & 3.50 & 2.60 &  7.6 &  99 & U028N2 & F. Marchis     \\
10 & 2017-05-05 & VLT   & SPHERE/IFS    &  1 & 2.09 & 1.44 & 25.6 & 179 & 099.D-0098 & J.-L. Beuzit   \\
11 & 2017-05-20 & VLT   & SPHERE/ZIMPOL &  1 & 2.06 & 1.56 & 28.3 & 164 & 199.C-0074 & P. Vernazza    \\
12 & 2018-08-05 & VLT   & SPHERE/ZIMPOL &  3 & 2.80 & 1.90 & 11.7 & 136 & 199.C-0074 & P. Vernazza    \\
    \hline
  \end{tabular}
\end{center}
\end{table*}

\begin{figure*}[ht]
    \includegraphics[width=0.8\textwidth]{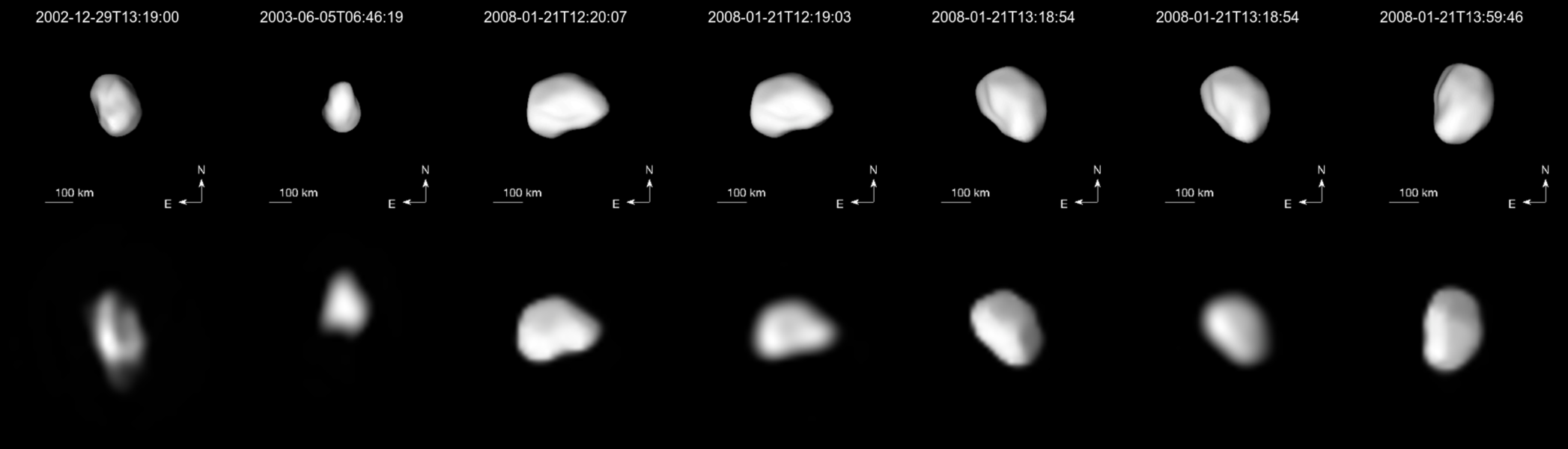}\\
    \includegraphics[width=0.8\textwidth]{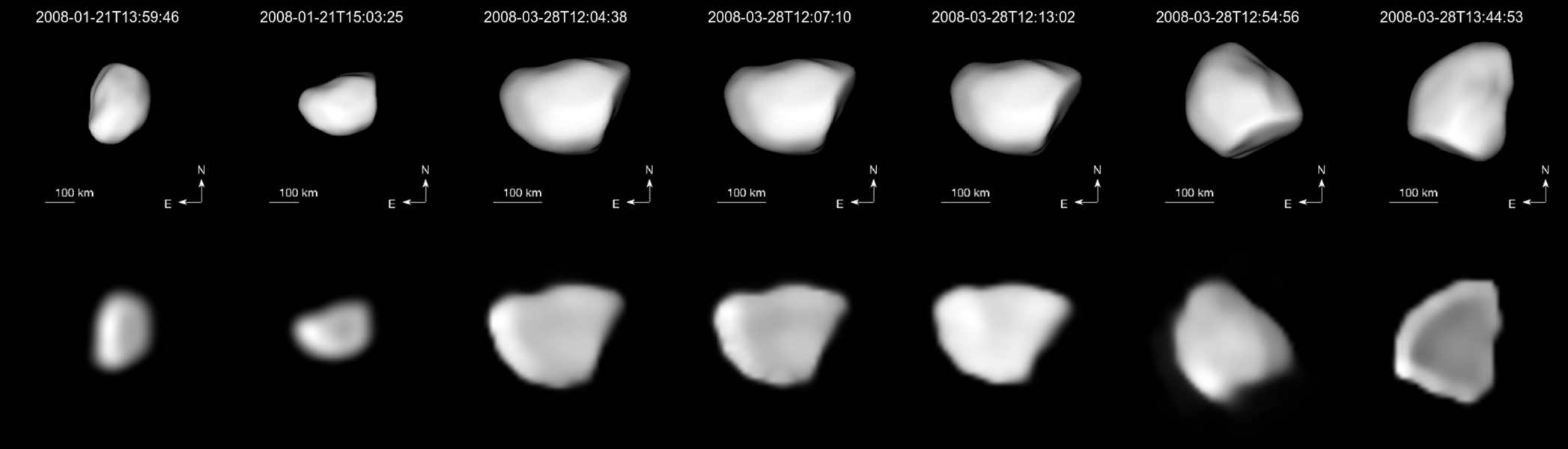}\\
    \includegraphics[width=0.8\textwidth]{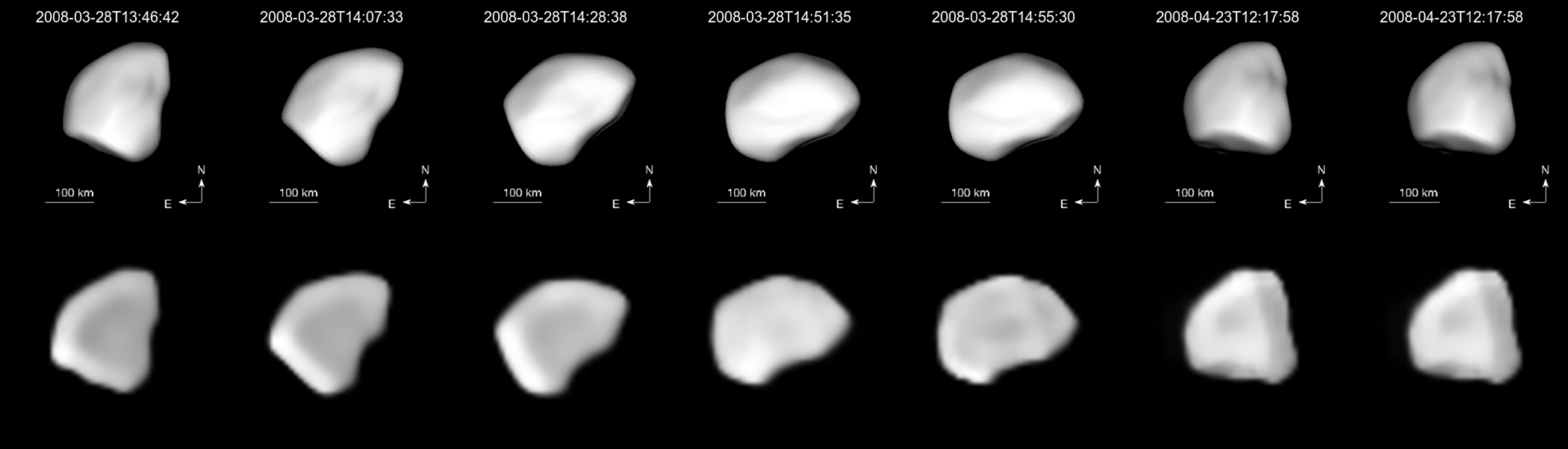}\\
    \includegraphics[width=0.8\textwidth]{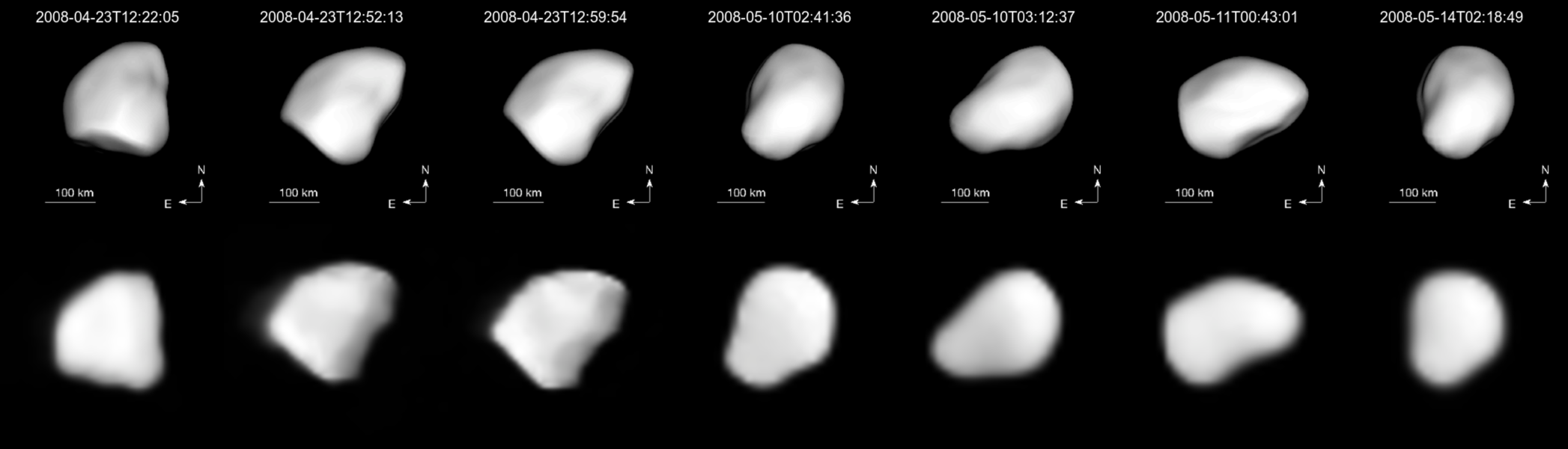}\\
    \includegraphics[width=0.6857\textwidth]{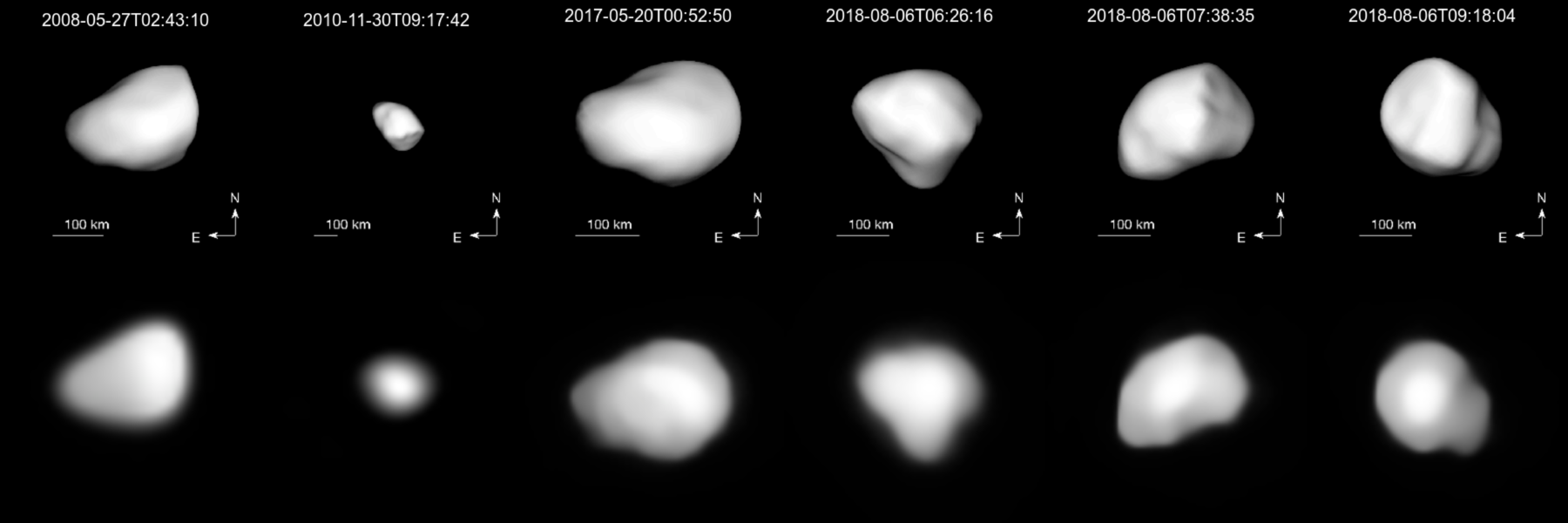}\\
  \begin{center}
    \caption[Images of Daphne]{
      Comparison of the shape
      model (top image of each row), oriented and projected on the
      plane of the sky at the 
      epoch of each disk-resolved observation (bottom image of each row).
      The brighter ring presents in some images \add{(e.g., second
        row, last column)} is an artifact from
      deconvolution.}
    \label{fig:ao}
  \end{center}
\end{figure*}

\begin{table*}
\begin{center}
  \caption{
    Date, number of positive and negative chords (\#$_p$ and \#$_n$), and
    average uncertainty in seconds ($\sigma_s$) and kilometers
    ($\sigma_{km}$) for each stellar occultation. 
    \label{tab:occ}
  }
  \begin{tabular}{rccrrrr}
    \hline\hline
     & Date & UT & \multicolumn{1}{c}{\#$_p$} & \multicolumn{1}{c}{\#$_n$} & 
     \multicolumn{1}{c}{$\sigma_s$} &
     \multicolumn{1}{c}{$\sigma_{km}$} \\
    && (h) &&& \multicolumn{1}{c}{(s)} & \multicolumn{1}{c}{(km)} \\
    \hline
    1   & 1999-07-02 & 20:27 &  21 &   3 &     1.21 &   46.150 \\
    2   & 2008-04-01 & 18:12 &   2 &   1 &     0.50 &    4.814 \\
    3   & 2012-01-09 & 10:39 &   1 &   3 &     1.00 &    3.224 \\
    4   & 2012-02-23 & 20:45 &   3 &   1 &     0.10 &    1.492 \\
    5   & 2012-03-02 & 18:49 &   1 &   4 &     2.00 &    3.411 \\
    6   & 2013-03-30 & 18:29 &   2 &   0 &     0.01 &    0.340 \\
    7   & 2013-09-05 & 22:26 &   4 &   2 &     0.44 &    7.655 \\
    8   & 2013-11-29 & 19:16 &   2 &   0 &     0.26 &   16.458 \\
    9   & 2016-01-17 & 22:42 &  19 &   0 &     0.41 & 1104.226 \\
    \hline
  \end{tabular}
\end{center}
\end{table*}

\begin{figure*}[ht]
  \includegraphics[width=\textwidth]{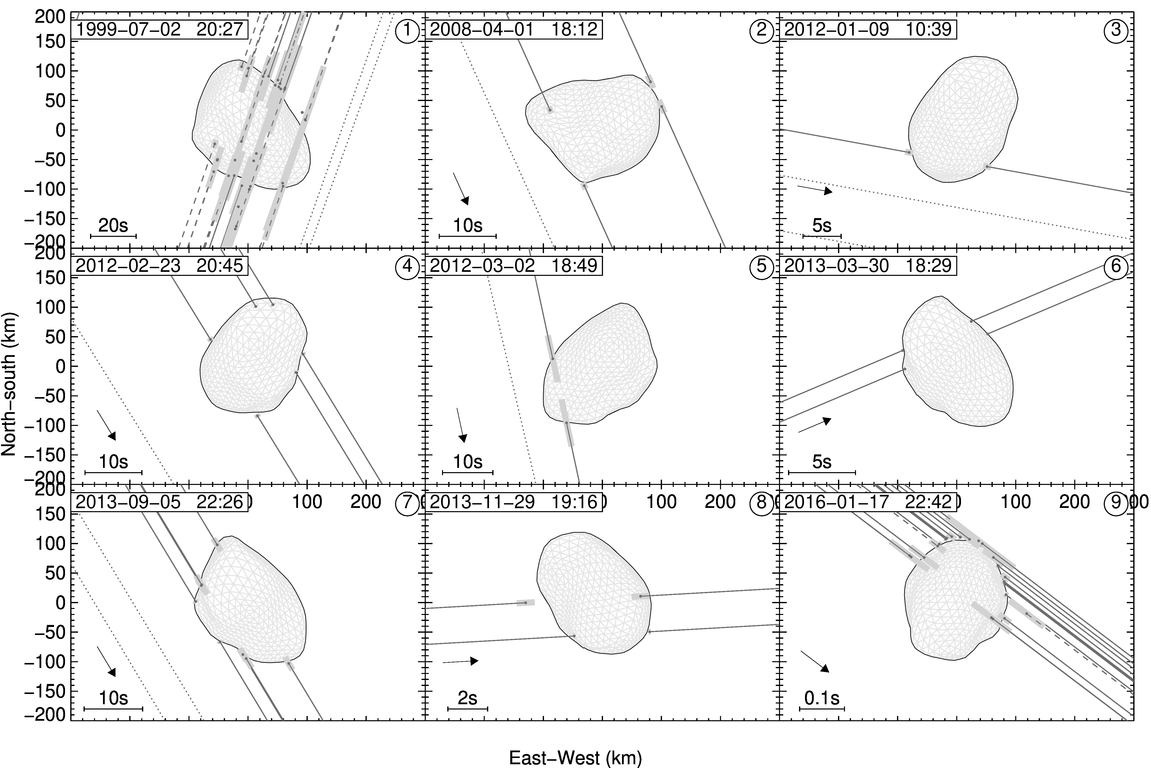}
  \caption[Stellar occultations by Daphne]{
    The \numb{eight} stellar occultations by Daphne, compared with the shape
    model (profile in black and facets in light gray)
    projected on the plane of the sky at the time of the
    occultation.
    Negative chords are represented by dotted lines, 
    visual timings by dashed lines, and 
    electronic timings by solid lines.
    Disappearance and reappearances timings are marked by filled
    circles, and their uncertainty by gray rectangles.
  }
  \label{fig:occ}
\end{figure*}


  \onecolumn
\begin{center}
  \begin{longtable}{rcrrrlrl}
    \caption{
      Date, duration ($\mathcal{L}$, in hours), number of points ($\mathcal{N}_p$), phase angle ($\alpha$), 
      filter, residual (against the shape model), and observers, for each 
      lightcurve. \label{tab:lc}
    }\\

    \hline\hline
    & Date & \multicolumn{1}{c}{$\mathcal{L}$} & \multicolumn{1}{c}{$\mathcal{N}_p$} &
    \multicolumn{1}{c}{$\alpha$} & \multicolumn{1}{c}{Filter} & \multicolumn{1}{c}{RMS} &
    \multicolumn{1}{c}{Observers} \\
    && \multicolumn{1}{c}{(h)} && \multicolumn{1}{c}{(\degr)}&& \multicolumn{1}{c}{(mag)} \\
    \hline
    \endfirsthead

    \multicolumn{8}{c}{{\tablename\ \thetable{} -- continued from previous page}} \\ 
    \hline\hline
    & Date & \multicolumn{1}{c}{$\mathcal{L}$} & \multicolumn{1}{c}{$\mathcal{N}_p$} &
    \multicolumn{1}{c}{$\alpha$} & \multicolumn{1}{c}{Filter} & \multicolumn{1}{c}{RMS} &
    \multicolumn{1}{c}{Observers} \\
    && \multicolumn{1}{c}{(h)} && \multicolumn{1}{c}{(\degr)}&& \multicolumn{1}{c}{(mag)} \\
    \hline
    \endhead

    \hline \multicolumn{8}{r}{{Continued on next page}} \\ \hline
    \endfoot

    \hline
    \endlastfoot

      1 & 1976-04-26 &  6.2 & 188 &  18.4 & V     &  0.024 & \citet{1977-AA-56-Scaltriti}   \\
      2 & 1976-05-02 &  6.1 & 163 &  16.8 & V     &  0.028 & \citet{1977-AA-56-Scaltriti}   \\
      3 & 1976-05-31 &  3.8 & 130 &  13.8 & V     &  0.029 & \citet{1977-AA-56-Scaltriti}   \\
      4 & 1981-06-16 &  3.7 &  14 &  21.1 & V     &  0.018 & \citet{1987-Icarus-70-Weidenschilling} \\
      5 & 1981-06-17 &  1.6 &   7 &  20.9 & V     &  0.018 & \citet{1987-Icarus-70-Weidenschilling} \\
      6 & 1981-07-25 &  4.2 &  34 &   9.8 & V     &  0.029 & \citet{1983-AA-54-Barucci} \\
      7 & 1981-07-23 &  4.3 &  58 &   9.8 & V     &  0.031 & \citet{1983-AA-54-Barucci} \\
      8 & 1981-08-03 &  4.8 &  78 &   7.4 & V     &  0.022 & \citet{1983-AA-54-Barucci} \\
      9 & 1981-08-04 &  6.0 &  64 &   7.4 & V     &  0.033 & \citet{1983-AA-54-Barucci} \\
     10 & 1981-08-06 &  5.6 &  23 &   7.2 & V     &  0.030 & \citet{1987-Icarus-70-Weidenschilling} \\
     11 & 1981-11-05 &  3.6 &  20 &  20.6 & V     &  0.050 & \citet{1987-Icarus-70-Weidenschilling} \\
     12 & 1981-12-01 &  1.3 &   7 &  18.8 & V     &  0.054 & \citet{1987-Icarus-70-Weidenschilling} \\
     13 & 1981-12-02 &  5.9 &  12 &  18.7 & V     &  0.028 & \citet{1987-Icarus-70-Weidenschilling} \\
     14 & 1982-09-29 &  4.8 &  20 &   7.5 & V     &  0.017 & \citet{1987-Icarus-70-Weidenschilling} \\
     15 & 1982-10-26 &  4.8 &  35 &   3.7 & V     &  0.033 & \citet{1983-AA-54-Barucci} \\
     16 & 1983-10-11 &  1.0 &   5 &  16.7 & V     &  0.035 & \citet{1987-Icarus-70-Weidenschilling} \\
     17 & 1983-10-15 &  1.8 &   7 &  16.4 & V     &  0.017 & \citet{1987-Icarus-70-Weidenschilling} \\
     18 & 1983-11-12 &  4.1 &  24 &  12.2 & V     &  0.015 & \citet{1987-Icarus-70-Weidenschilling} \\
     19 & 1983-11-14 &  5.8 &  14 &  11.8 & V     &  0.024 & \citet{1987-Icarus-70-Weidenschilling} \\
     20 & 1983-12-28 &  6.0 &  83 &   8.1 & V     &  0.068 & \citet{1983-AA-54-Barucci} \\
     21 & 1983-12-29 &  0.8 &  15 &   8.2 & V     &  0.032 & \citet{1983-AA-54-Barucci} \\
     22 & 1983-12-30 &  4.0 &  77 &   8.4 & V     &  0.047 & \citet{1983-AA-54-Barucci} \\
     23 & 1984-02-21 &  4.8 &  34 &  18.1 & V     &  0.028 & \citet{1987-Icarus-70-Weidenschilling} \\
     24 & 1985-01-18 &  2.2 &  21 &  25.8 & V     &  0.020 & \citet{1987-Icarus-70-Weidenschilling} \\
     25 & 1985-01-19 &  3.9 &  21 &  25.7 & V     &  0.023 & \citet{1987-Icarus-70-Weidenschilling} \\
     26 & 1985-04-11 &  5.1 &  13 &   5.8 & V     &  0.014 & \citet{1987-Icarus-70-Weidenschilling} \\
     27 & 1987-10-17 &  5.8 &  25 &   8.7 & V     &  0.019 & \citet{1987-Icarus-70-Weidenschilling} \\
     28 & 1988-12-20 &  5.9 &  37 &  11.5 & V     &  0.042 & \citet{1987-Icarus-70-Weidenschilling} \\
     29 & 1988-12-21 &  5.8 &  15 &  11.2 & V     &  0.035 & \citet{1987-Icarus-70-Weidenschilling} \\
     30 & 2001-11-21 &  7.2 & 103 &   5.4 & V     &  0.023 & L.~Bernasconi   \\
     31 & 2001-11-22 &  8.4 & 107 &   5.4 & V     &  0.023 & L.~Bernasconi   \\
     32 & 2001-11-23 &  7.0 & 102 &   5.5 & V     &  0.027 & L.~Bernasconi   \\
     33 & 2001-11-24 &  8.8 &  87 &   5.5 & V     &  0.022 & L.~Bernasconi   \\
     34 & 2008-06-25 &  2.3 &  49 &  12.9 & clear &  0.018 & SuperWASP - J.~Grice   \\
     35 & 2009-10-16 &  3.0 &  45 &   4.3 & clear &  0.021 & SuperWASP - J.~Grice   \\
     36 & 2009-09-25 &  2.7 & 117 &   8.2 & V     &  0.019 & ASTEP          \\
     37 & 2009-10-16 &  2.8 &  34 &  11.2 & clear &  0.015 & SuperWASP - J.~Grice   \\
     38 & 2017-01-08 &  2.3 &  31 &  22.0 & R     &  0.024 & Vachier, Klotz, Teng, Peyrot, Thierry, Berthier\\
     39 & 2018-07-09 & 10.6 & 651 &  19.1 & R     &  0.023 & E.~Jehin                        \\
     40 & 2018-07-10 &  6.0 & 507 &  18.9 & R     &  0.024 & E.~Jehin                        \\
     41 & 2018-08-03 &  3.6 & 200 &  12.4 & R     &  0.023 & S.~Fauvaud \\                            
     42 & 2018-08-04 &  6.2 & 200 &  12.1 & R     &  0.020 & S.~Fauvaud \\                           
     43 & 2018-08-05 &  2.1 & 200 &  11.7 & R     &  0.019 & S.~Fauvaud \\                           
     44 & 2018-08-05 &  1.9 & 137 &  11.4 & R     &  0.013 & S.~Fauvaud \\                           
     45 & 2018-08-06 &  1.8 & 136 &  11.1 & R     &  0.012 & S.~Fauvaud \\                           
     46 & 2018-08-11 &  1.6 & 139 &   9.3 & R     &  0.009 & S.~Fauvaud \\                           
     47 & 2018-08-14 &  1.8 & 139 &   8.2 & R     &  0.008 & S.~Fauvaud \\                           
     48 & 2018-08-15 &  1.4 & 105 &   7.8 & R     &  0.007 & S.~Fauvaud \\
     49 & 2018-11-19 &  3.1 &  52 &  18.6 & R     &  0.016 & Vachier, Klotz, Teng, Peyrot, Thierry, Berthier\\
     \hline
  \end{longtable}
\end{center}
\twocolumn

\begin{figure*}[]
  \centering
  \begin{subfigure}{\textwidth}
    \includegraphics[width=.9\linewidth]{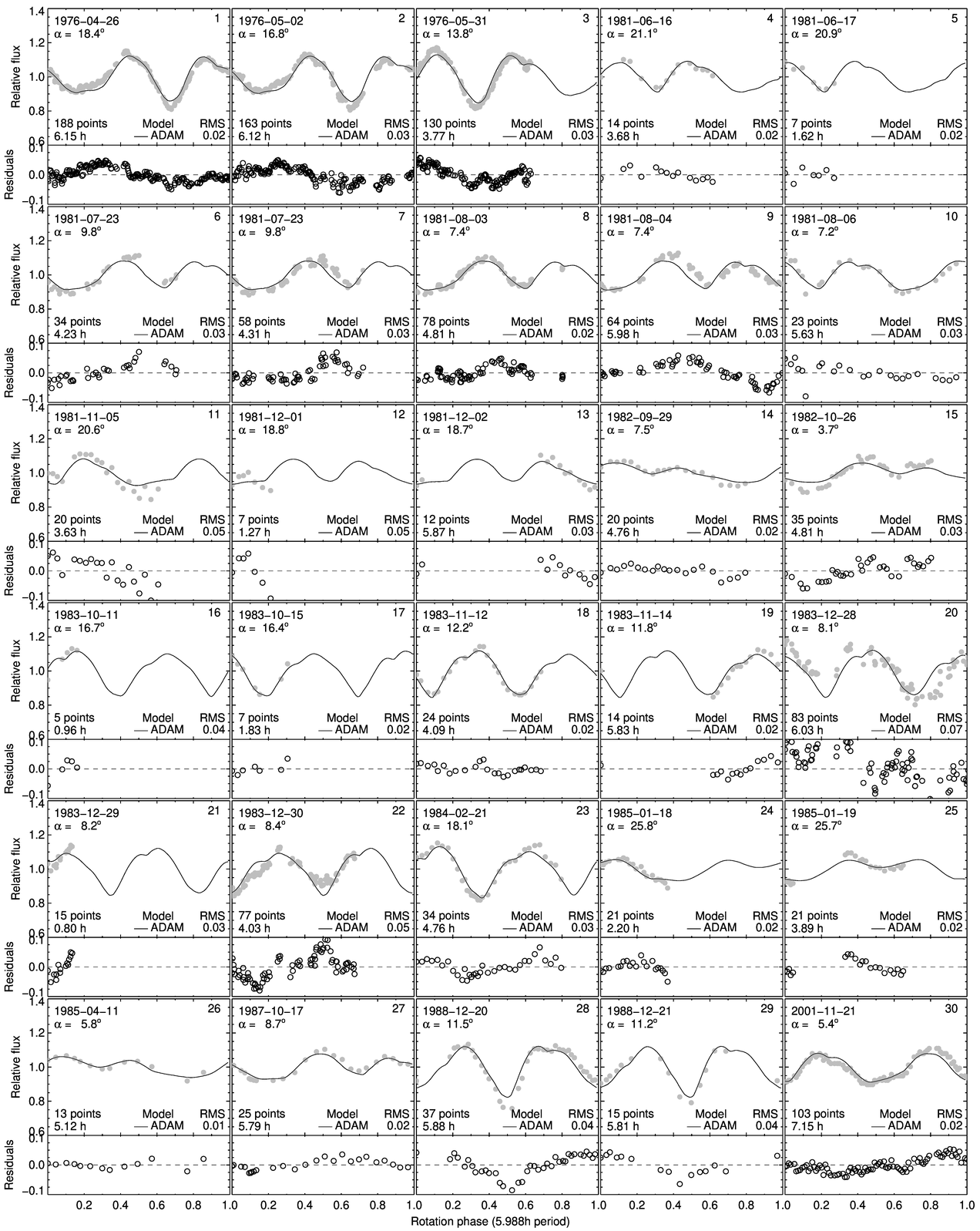}
  \end{subfigure}
  \caption[Optical lightcurves of Daphne]{
    The optical lightcurves of Daphne (grey spheres),
    compared with the synthetic lightcurves generated with the shape
    model (black lines). 
    On each panel, the observing date, number of points, duration
    of the lightcurve (in hours), and RMS residuals between the
    observations and the synthetic lightcurves
    are displayed.
    Measurement uncertainties are seldom provided by the
    observers but can be estimated from the spread of measurements. 
  \label{app:lc}}
\end{figure*}
\clearpage
\begin{figure*}[]
  \centering
  \ContinuedFloat
  \captionsetup{list=off}
  \begin{subfigure}{.9\textwidth}
    \includegraphics[width=\linewidth]{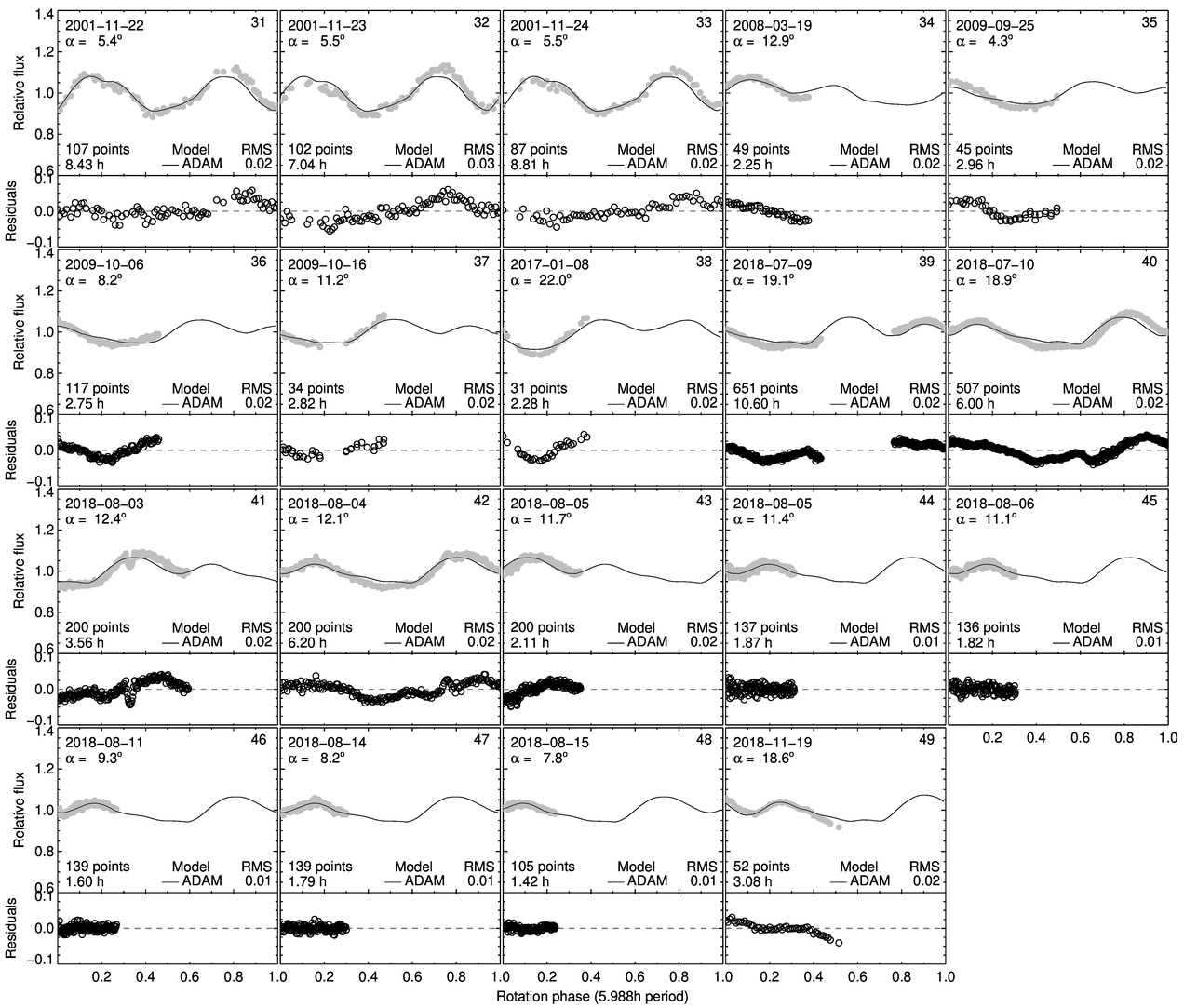}
  \end{subfigure}
  \caption{Cont'd}
\end{figure*}

\clearpage
\section{Magnitude and position of the satellite\label{app:dyn}}

\begin{table*}
\begin{center}
  \caption[Astrometry of the satellite of Daphne]{Astrometry of the
    satellite of Daphne.
    Date, mid-observing time (UTC), telescope, camera, filter, 
    astrometry ($X$
    is aligned with Right Ascension, and $Y$ with Declination, and
    $o$ and $c$ indices stand for observed and computed positions),
    and photometry (magnitude difference $\Delta M$ with uncertainty $\delta M$).
    \label{tab:genoid}
  }
  \begin{tabular}{cclllrrrrrrr}
    \hline\hline
     Date & UTC & Tel. & Cam. & Filter &
     \multicolumn{1}{c}{$X_o$} &
     \multicolumn{1}{c}{$Y_o$} &
     \multicolumn{1}{c}{$X_{o-c}$} &
     \multicolumn{1}{c}{$Y_{o-c}$} &
     \multicolumn{1}{c}{$\sigma$} &
     \multicolumn{1}{c}{$\Delta M$} &
     \multicolumn{1}{c}{$\delta M$} \\
    &&&&& 
     \multicolumn{1}{c}{(mas)} & \multicolumn{1}{c}{(mas)} &
     \multicolumn{1}{c}{(mas)} & \multicolumn{1}{c}{(mas)} & 
     \multicolumn{1}{c}{(mas)} & 
     \multicolumn{1}{c}{(mag)} &\multicolumn{1}{c}{(mag)}  \\
    \hline
2008-03-28 & 12:05:30.1 & Keck     & NIRC2    & J        & -554 &   60 &   -6 &   -5 &   9.94 &  -9.59 &   0.85 \\
2008-03-28 & 12:07:59.5 & Keck     & NIRC2    & J        & -549 &   53 &   -1 &   -7 &   9.94 & -10.06 &   0.64 \\
2008-03-28 & 12:56:28.9 & Keck     & NIRC2    & H        & -531 &  -19 &   -5 &    5 &   9.94 &   --   &   --   \\
2008-03-28 & 13:45:16.2 & Keck     & NIRC2    & H        & -476 & -110 &    8 &    0 &   9.94 &   --   &   --   \\
2008-03-28 & 13:48:16.8 & Keck     & NIRC2    & H        & -475 & -115 &    6 &    1 &   9.94 &  -9.70 &   0.14 \\
2008-03-28 & 14:09:03.8 & Keck     & NIRC2    & H        & -457 & -160 &    1 &   -7 &   9.94 &  -9.10 &   2.19 \\
2008-03-28 & 14:30:11.3 & Keck     & NIRC2    & H        & -435 & -187 &   -3 &    0 &   9.94 &   --   &   --   \\
2008-03-28 & 14:53:08.0 & Keck     & NIRC2    & Kp       & -403 & -215 &   -3 &    9 &   9.94 &  -9.84 &   0.73 \\
2008-03-28 & 14:57:06.5 & Keck     & NIRC2    & Kp       & -397 & -222 &   -3 &    7 &   9.94 &   --   &   --   \\
2008-03-28 & 15:00:34.9 & Keck     & NIRC2    & H        & -389 & -229 &    0 &    6 &   9.94 &   --   &   --   \\
2008-04-23 & 12:17:26.4 & Keck     & NIRC2    & H        & -424 &  376 &  -16 &    0 &   9.94 &   --   &   --   \\
2008-04-23 & 12:53:37.9 & Keck     & NIRC2    & H        & -453 &  361 &    6 &   19 &   9.94 &   --   &   --   \\
2008-04-23 & 12:58:51.4 & Keck     & NIRC2    & H        & -452 &  347 &   14 &   11 &   9.94 &   --   &   --   \\
2008-05-10 & 02:43:04.5 & VLT      & NACO     & H        &  550 & -177 &    3 &   -1 &  13.24 &  -9.56 &   0.78 \\
2008-05-10 & 02:47:42.2 & VLT      & NACO     & H        &  552 & -167 &    3 &    1 &  13.24 &  -9.53 &   1.49 \\
2008-05-10 & 03:14:05.9 & VLT      & NACO     & H        &  559 & -124 &    4 &    9 &  13.24 &  -9.52 &   0.84 \\
2008-05-10 & 03:18:44.1 & VLT      & NACO     & H        &  556 & -112 &    1 &   14 &  13.24 &  -9.29 &   1.13 \\
2008-05-14 & 02:20:19.1 & VLT      & NACO     & H        & -516 &  142 &   11 &  -24 &  13.24 &  -9.28 &   0.37 \\
2008-05-14 & 02:24:59.4 & VLT      & NACO     & H        & -523 &  139 &    6 &  -21 &  13.24 &  -9.02 &   0.19 \\
2008-05-27 & 02:44:38.8 & VLT      & NACO     & H        &  449 & -256 &   11 &   -1 &  13.24 &   --   &   --   \\
2008-05-27 & 02:49:12.7 & VLT      & NACO     & H        &  445 & -251 &    2 &   -1 &  13.24 &  -9.68 &   0.65 \\
2017-05-05 & 22:51:25.1 & VLT      & IFS      & YJH      &  404 &  158 &   -6 &  -10 &  10.00 &   --   &   --   \\
2017-05-20 & 01:01:27.4 & VLT      & ZIMPOL   & R        & -360 &   35 &    9 &    8 &  10.00 &   --   &   --   \\
2018-08-06 & 06:28:30.9 & VLT      & ZIMPOL   & R        & -268 &  208 &   -5 &    4 &  10.00 &   --   &   --   \\
2018-08-06 & 06:33:09.9 & VLT      & ZIMPOL   & R        & -266 &  203 &   -6 &   -3 &  10.00 &   --   &   --   \\
2018-08-06 & 07:40:49.4 & VLT      & ZIMPOL   & R        & -193 &  244 &    7 &   -1 &  10.00 &   --   &   --   \\
2018-08-06 & 07:45:25.2 & VLT      & ZIMPOL   & R        & -186 &  249 &    9 &    0 &  10.00 &   --   &   --   \\
2018-08-06 & 07:59:15.8 & VLT      & ZIMPOL   & R        & -169 &  251 &   13 &   -2 &  10.00 &  -9.97 &   0.63 \\
2018-08-06 & 09:20:18.1 & VLT      & ZIMPOL   & R        &  -92 &  271 &   -1 &   -2 &  10.00 &   --   &   --   \\
2018-08-06 & 09:24:57.2 & VLT      & ZIMPOL   & R        &  -87 &  272 &   -1 &   -1 &  10.00 & -10.02 &   1.34 \\
\hline
&&&&& \multicolumn{3}{c}{Average}                 2 &    0 &   15 &   -9.49 &   0.28 \\ 
&&&&& \multicolumn{3}{c}{Standard deviation}      7 &    9 &    2 &    0.32 &   0.18 \\
    \hline
  \end{tabular}
\end{center}
\end{table*}

\begin{figure}[ht]
  \begin{center}
    \includegraphics[width=\hsize]{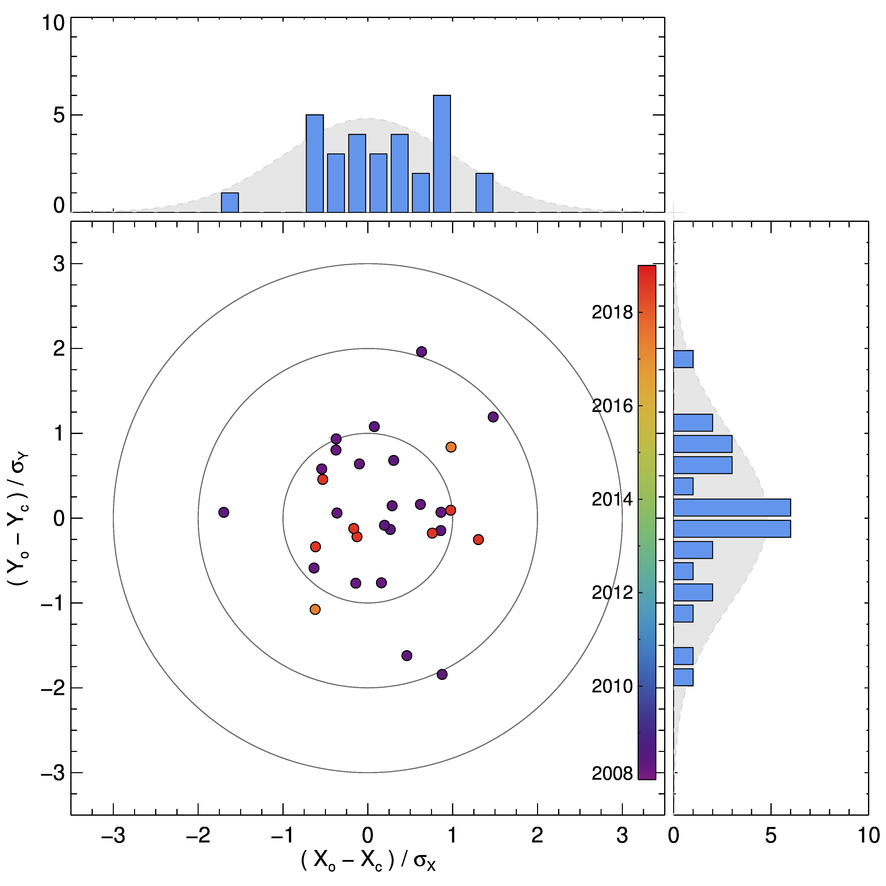}
    \caption[Residuals on the orbit determination]{
      Distribution of residuals for the satellite between the
      observed (index o)
      and
      predicted (index c) positions, normalized by the uncertainty on
      the measured 
      positions ($\sigma$), and color-coded by observing epoch.
      X stands for right ascension and Y for declination.
      The three large gray circles represent the 1, 2, and 3 $\sigma$
      limits \add{(typically 10\,mas at 1\,$\sigma$)}. 
      The top panel shows the histogram of residuals along X, and the
      right panel the residuals along Y. The light gray Gaussian in the
      background has a standard deviation of one.
    }
    \label{fig:rmsorb}
  \end{center}
\end{figure}

\begin{figure}[ht]
  \includegraphics[width=\hsize]{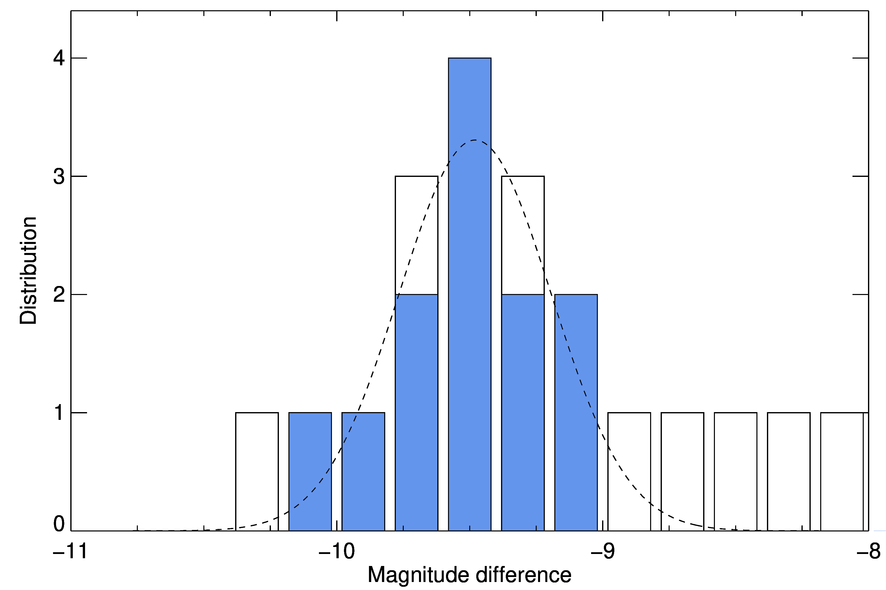}
  \caption[Magnitude of the satellite with respect to Daphne]{
    Distribution of the magnitude differences between Daphne and its
    satellite. The open bars represent all measurements, and the blue
    bars those more precise than 0.75 magnitude.
    The dashed black line represents the normal distribution fit to our
    results, with a mean and standard deviation of \numb{\DMag}.
  }
  \label{fig:dmag}
\end{figure}

\clearpage
\section{Mass and diameter of Ch and Cgh asteroids\label{app:cgh}}

\longtab[1]{
\begin{longtable}{rlllccrrcc}
\caption{\label{tab:ssodd}Dynamical class (IMB, MMB, and OMB  stand
  for Inner, Middle, and Outer belt respectively), taxonomic class, 
  mass ($\mathcal{M}$), diameter ($\mathcal{D}$), and density ($\rho$)
  of Ch/Cgh asteroids used in this work, compiled following the
  recipes described in \citet{2012-PSS-73-Carry}.
  Individual diameter and mass estimates are listed in
  Tables~\ref{tab:alldiam} and~\ref{tab:allmass}.
}\\ 
\hline\hline
  \multicolumn{1}{c}{\#} & \multicolumn{1}{c}{Name} & 
  \multicolumn{1}{c}{Dyn.} & \multicolumn{1}{c}{Taxo.}  &
  \multicolumn{1}{c}{$\mathcal{M}$} & \multicolumn{1}{c}{$\sigma_\mathcal{M}$} &
  \multicolumn{1}{c}{$\mathcal{D}$} & \multicolumn{1}{c}{$\sigma_\mathcal{D}$} &
  \multicolumn{1}{c}{$\rho$} & \multicolumn{1}{c}{$\sigma_\rho$} \\
  &&&&\multicolumn{1}{c}{(kg)} & \multicolumn{1}{c}{(kg)} &
  \multicolumn{1}{c}{(km)} & \multicolumn{1}{c}{(km)} &
  \multicolumn{1}{c}{(g$\cdot$cm$^{-3}$)} & \multicolumn{1}{c}{(g$\cdot$cm$^{-3}$)} \\
\hline
\endfirsthead
\caption{continued.}\\
\hline\hline
  \multicolumn{1}{c}{\#} & \multicolumn{1}{c}{Name} & 
  \multicolumn{1}{c}{Dyn.} & \multicolumn{1}{c}{Taxo.}  &
  \multicolumn{1}{c}{$\mathcal{M}$} & \multicolumn{1}{c}{$\delta \mathcal{M}$} &
  \multicolumn{1}{c}{$\mathcal{D}$} & \multicolumn{1}{c}{$\delta \mathcal{D}$} &
  \multicolumn{1}{c}{$\rho$} & \multicolumn{1}{c}{$\delta \rho$} \\
  &&&&\multicolumn{1}{c}{(kg)} & \multicolumn{1}{c}{(kg)} &
  \multicolumn{1}{c}{(km)} & \multicolumn{1}{c}{(km)} &
  \multicolumn{1}{c}{(g$\cdot$cm$^{-3}$)} & \multicolumn{1}{c}{(g$\cdot$cm$^{-3}$)} \\
\hline
\endhead
\hline
\endfoot
    13&Egeria        &MMB      &Ch  & 8.42$\cdot 10^{18}$&2.28$\cdot 10^{18}$&206.56&   8.22&  1.82&  0.54 \\
    19&Fortuna       &IMB      &Ch  & 9.03$\cdot 10^{18}$&7.13$\cdot 10^{17}$&207.54&   5.30&  1.93&  0.21 \\
    34&Circe         &MMB      &Ch  & 4.16$\cdot 10^{18}$&6.27$\cdot 10^{17}$&114.34&   3.23&  5.31&  0.92 \\
    38&Leda          &MMB      &Cgh & 3.18$\cdot 10^{18}$&5.97$\cdot 10^{17}$&118.62&   4.22&  3.64&  0.79 \\
    41&Daphne        &MMB      &Ch  & 6.10$\cdot 10^{18}$&7.50$\cdot 10^{17}$&187.00&   7.50&  1.77&  0.22 \\
    48&Doris         &OMB      &Ch  & 7.71$\cdot 10^{18}$&1.57$\cdot 10^{18}$&208.37&  12.02&  1.63&  0.44 \\
    49&Pales         &OMB      &Ch  & 4.10$\cdot 10^{18}$&1.77$\cdot 10^{18}$&149.97&   3.78&  2.32&  1.02 \\
    50&Virginia      &MMB      &Ch  & 6.26$\cdot 10^{17}$&2.06$\cdot 10^{17}$& 88.51&   4.65&  1.72&  0.63 \\
\end{longtable}
}

\longtab[2]{
\begin{longtable}{rlrrll}
\caption{\label{tab:alldiam}
  The diameter estimates ($\mathcal{D}$) for all the Ch/Cgh asteroids available in the literature.
    For each, the 3\,$\sigma$ uncertainty, method, and 
    bibliographic reference are reported. The methods are
    \textsc{im-te}: Ellipsoid from Imaging,
    \textsc{lcimg}: 3-D Model scaled with Imaging, 
    \textsc{adam} \& \textsc{koala}: Multidata 3-D Modeling,
    \textsc{occ}: Stellar Occultations,
    \textsc{stm}: Standard Thermal Model, 
    \textsc{neatm}: Near-Earth Asteroid Thermal Model, and
    \textsc{tpm}: Thermophysical Model.
}\\ 
\hline\hline
     \multicolumn{1}{c}{\#} & \multicolumn{1}{c}{Name} & 
     \multicolumn{1}{c}{$\mathcal{D}$} & \multicolumn{1}{c}{$\delta \mathcal{D}$} &
     \multicolumn{1}{c}{Method} & \multicolumn{1}{c}{Reference}  \\
    & \multicolumn{1}{c}{(km)} & \multicolumn{1}{c}{(km)} \\
\hline
\endfirsthead
\caption{continued.}\\
\hline\hline
     \multicolumn{1}{c}{\#} & \multicolumn{1}{c}{Name} & 
     \multicolumn{1}{c}{$\mathcal{D}$} & \multicolumn{1}{c}{$\delta \mathcal{D}$} &
     \multicolumn{1}{c}{Method} & \multicolumn{1}{c}{Reference}  \\
    & \multicolumn{1}{c}{(km)} & \multicolumn{1}{c}{(km)} \\
\hline
\endhead
\hline
\endfoot
    13 & Egeria          & 244.0 &  73.2 &\textsc{stm}   &\citet{PDSSBN-TRIAD}                 \\
    13 & Egeria          & 207.6 &  24.9 &\textsc{stm}   &\citet{PDSSBN-IRAS}                  \\
    13 & Egeria          & 203.1 &   5.9 &\textsc{occ}   &\citet{PDSSBN-OCC}                   \\
    13 & Egeria          & 223.1 &  10.4 &\textsc{stm}   &\citet{2010-AJ-140-Ryan}             \\
    13 & Egeria          & 226.1 &  28.5 &\textsc{neatm} &\citet{2010-AJ-140-Ryan}             \\
    13 & Egeria          & 203.4 &   7.7 &\textsc{stm}   &\citet{2011-PASJ-63-Usui}            \\
    13 & Egeria          & 227.0 &  77.9 &\textsc{neatm} &\citet{2011-ApJ-741-Masiero}         \\
    13 & Egeria          & 202.6 & 150.2 &\textsc{neatm} &\citet{2015-ApJ-814-Nugent}          \\
\hline
\end{longtable}}

\longtab[3]{
\begin{longtable}{rlrrll}
\caption{\label{tab:allmass}
  The mass ($\mathcal{M}$) for all the Ch/Cgh asteroids available in the literature.
    For each, the 3\,$\sigma$ uncertainty, method, and 
    bibliographic reference are reported. The methods are
    \textsc{defl}: Deflection, 
    \textsc{ephem}: Ephemeris,
    \textsc{bimg}: Binary: Imaging, and
    \textsc{bgeno}: Binary: Genoid.}\\ 
\hline\hline
     \multicolumn{1}{c}{\#} & \multicolumn{1}{c}{Name} & 
     \multicolumn{1}{c}{$\mathcal{M}$} & \multicolumn{1}{c}{$\delta \mathcal{M}$} &
     \multicolumn{1}{c}{Method} & \multicolumn{1}{c}{Reference}  \\
    & \multicolumn{1}{c}{(kg)} & \multicolumn{1}{c}{(kg)} \\
\hline
\endfirsthead
\caption{continued.}\\
\hline\hline
     \multicolumn{1}{c}{\#} & \multicolumn{1}{c}{Name} & 
     \multicolumn{1}{c}{$\mathcal{M}$} & \multicolumn{1}{c}{$\delta \mathcal{M}$} &
     \multicolumn{1}{c}{Method} & \multicolumn{1}{c}{Reference}  \\
    & \multicolumn{1}{c}{(kg)} & \multicolumn{1}{c}{(kg)} \\
\hline
\endhead
\hline
\endfoot
    13 & Egeria          & 1.63 $\times 10^{19}$ & 9.54 $\times 10^{18}$ &\textsc{defl}  &\citet{2008-DPS-40-Baer}             \\
    13 & Egeria          & 6.17 $\times 10^{18}$ & 1.85 $\times 10^{18}$ &\textsc{ephem} &\citet{2009-SciNote-Folkner}         \\
    13 & Egeria          & 1.59 $\times 10^{19}$ & 1.31 $\times 10^{19}$ &\textsc{defl}  &\citet{2011-AJ-141-Baer}             \\
    13 & Egeria          & 1.29 $\times 10^{19}$ & 1.41 $\times 10^{19}$ &\textsc{defl}  &\citet{2011-AJ-142-Zielenbach}       \\
    13 & Egeria          & 7.39 $\times 10^{18}$ & 9.72 $\times 10^{18}$ &\textsc{defl}  &\citet{2011-AJ-142-Zielenbach}       \\
    13 & Egeria          & 6.07 $\times 10^{18}$ & 9.60 $\times 10^{18}$ &\textsc{defl}  &\citet{2011-AJ-142-Zielenbach}       \\
    13 & Egeria          & 8.26 $\times 10^{18}$ & 1.83 $\times 10^{19}$ &\textsc{defl}  &\citet{2011-AJ-142-Zielenbach}       \\
    13 & Egeria          & 9.37 $\times 10^{18}$ & 4.65 $\times 10^{18}$ &\textsc{ephem} &\citet{2012-SciNote-Fienga}          \\
\hline
\end{longtable}}

\end{document}